\documentclass[10pt]{article}
\usepackage{fullpage}
\usepackage{amsmath}
\usepackage{amssymb}
\usepackage{amsfonts}
\usepackage{comment}
\usepackage{color}
\usepackage{cite}
\usepackage{subcaption}
\usepackage{graphicx}
\usepackage{enumitem}

\def\##1{\underline{#1}}
\def\=#1{\underline{\underline{#1}}}

\def\+
#1{\underline{\bf #1}}
\def\*#1{\underline{\underline{\bf #1}}}

\def\r#1{(\ref{#1})}

\def\c#1{\cite{#1}}

\def\le{\left(}
\def\ri{\right)}
\def\les{\left[}
\def\ris{\right]}
\def\lec{\left\{}
\def\ric{\right\}}
\def\lek{[{\kern 0.1em}}
\def\rik{{\kern 0.1em}]}

\def\.{\mbox{ \tiny{$^\bullet$} }}

\def\eps{\varepsilon}

\def\epso{\eps_{\scriptscriptstyle 0}}
\def\lambdao{\lambda_{\scriptscriptstyle 0}}

\def\ux{\hat{\#u}_{\rm x}}
\def\uy{\hat{\#u}_{\rm y}}
\def\uz{\hat{\#u}_{\rm z}}

\def\un{\#{u}_{\rm n}}
\def\ut{\#{u}_\tau}
\def\ub{\#{u}_{\rm b}}

\def\degC{\,^\circ{\rm C}}

\begin{document}

\begin{center}

\LARGE{ {\bf   Towards Morphologically Induced Anisotropy in Thermally Hysteretic Dielectric
Properties of Vanadium Dioxide
}}
\end{center}
\begin{center}
\vspace{10mm} \large
 
 {Tom G. Mackay}\footnote{E--mail: T.Mackay@ed.ac.uk.}\\
{\em School of Mathematics and
   Maxwell Institute for Mathematical Sciences\\
University of Edinburgh, Edinburgh EH9 3FD, UK}\\
and\\
 {\em NanoMM~---~Nanoengineered Metamaterials Group\\ Department of Engineering Science and Mechanics\\
Pennsylvania State University, University Park, PA 16802--6812,
USA}
 \vspace{3mm}\\
 {Akhlesh  Lakhtakia}\\
 {\em NanoM~---~Nanoengineered Metamaterials Group\\ Department of Engineering Science and Mechanics\\
Pennsylvania State University, University Park, PA 16802--6812, USA}

\normalsize

\end{center}

\begin{center}
\vspace{5mm} {\bf Abstract}
\end{center}

The Bruggeman homogenization formalism was used to numerically investigate the dielectric properties of a columnar thin film (CTF) made from
vanadium dioxide. 
For visible and near-infrared wavelengths, the CTF is electromagnetically equivalent to a homogeneous orthorhombic material. 
Over the  $58\degC$--$72\degC$ temperature range,
the eigenvalues of
the CTF's relative permittivity dyadic are highly sensitive to temperature, and vary according to whether the CTF is being heated or cooled.
 The  anisotropy revealed  through the eigenvalues, and the anisotropy of the associated hysteresis, were investigated 
 in relation to temperature for
  CTFs of different porosities and  columnar cross sections.
When the free-space wavelength is 800~nm, the CTF is a dissipative dielectric material that exhibits temperature-dependent anisotropy and anisotropic hysteresis. In contrast, when
the free-space wavelength is 1550~nm, the CTF can be either a dissipative dielectric material, a hyperbolic material or a metal-like material, depending on the temperature and
   the porosity of the CTF.  As the porosity of the CTF   decreases from $0.55$ to $0.3$,
    the anisotropy of the CTF becomes  more pronounced, as does  
  the anisotropy of the hysteresis. Only relatively modest variations in anisotropy and hysteresis arise in response to varying the columnar cross-sectional shape, as compared to the variations induced by varying the porosity. 
 
 \vspace{5mm}
\vspace{5mm}
 
\section{Introduction}

Materials, metamaterials, and metasurfaces with thermally controllable dielectric properties are of interest for numerous applications involving
reconfigurable and multifunctional devices 
\c{Wang,Maguid,Huang,Lakh-multi}. 
 Vanadium dioxide (VO${}_2$) is especially interesting in this context as it undergoes a thermally-induced hysteretic phase change over the temperature range 
 $58\degC$--$72\degC$,
 in the visible, infrared, and terahertz spectral regimes 
\c{Dicken,Seo,Bonora,Cueff,Sereb2022,Lu,Shi}.

The electromagnetic response of
 VO${}_2$
 is isotropic,
 as encapsulated by its  (complex-valued) relative
  permittivity $\eps_{\text{VO}_2}$.
  However,
  the crystal structure of  VO${}_2$ is not
cubic~---~instead, it
 is monoclinic at temperatures below $58\degC$ and 
  tetragonal at temperatures above  $72\degC$ \c{Morin}.
  Both monoclinic and tetragonal crystals coexist in varying proportions in the $58\degC$--$72\degC$ temperature range.
 Furthermore, the value of $\eps_{\text{VO}_2}$ 
  for monoclinic   
   VO${}_2$ is significantly different to its
  value for tetragonal
  VO${}_2$,  at a given free-space wavelength  $\lambdao$.
  
  The transition from monoclinic VO${}_2$  to tetragonal VO${}_2$  can be achieved by heating and it can be reversed by cooling. 
  And this transition is
hysteretic. That is, 
at a temperature in the 
$58\degC$--$72\degC$ range, 
the value of 
$\eps_{\text{VO}_2}$  depends on whether VO${}_2$ was heated up or cooled down to reach that temperature \c{Cormier}. Since VO${}_2$ is
hysteretically biphasic,
$\eps_{\text{VO}_2}$  may be regarded as a double-valued function of temperature; hence, we write $\eps_{\text{VO}_2} = \eps^{\rm{heat}}_{\text{VO}_2}$ 
for the heating phase and $\eps_{\text{VO}_2} = \eps^{\rm{cool}}_{\text{VO}_2}$ 
for the cooling phase.

The prospect of  an anisotropic material, metamaterial or metasurface whose  dielectric properties for
 propagation in
 different directions  
  are  thermally controllable
 presents opportunities for  directional control and tuning in applications involving reconfigurable and multifunctional devices.
How could such a  prospect be realized? A practical means of realizing thermal directional control is to use VO${}_2$  to fabricate a columnar thin film (CTF) \c{HW,STF_Book,Ibrahim}. 
A CTF is a forest of parallel columns, inclined at a certain angle to a substrate upon which they are deposited by a technique such as  oblique angle deposition.
Macroscopically, a CTF is electromagnetically equivalent to a homogeneous anisotropic material with orthorhombic symmetry. 

 In the following sections, the Bruggeman homogenization formalism \c{Kanaun,Ward,MAEH} is employed to estimate the relative permittivity parameters of  a CTF made from VO${}_2$, and the effects of
 temperature on anisotropy and on the anisotropy of hysteresis are thereby investigated.
The following notation is adopted: 3-vectors are underlined once and 3$\times$3 dyadics are underlined twice. The triad of Cartesian basis vectors
is denoted as $\lec \ux, \uy, \uz \ric$.

\section{Homogenization model of columnar thin film}

\subsection{Preliminaries}

A bulk material is evaporated and a collimated vapor flux is directed in a low-pressure
chamber towards  a planar substrate to
form a CTF \c{HW,STF_Book,Gauntt,Vepa}.  Suppose that the  planar substrate is
 parallel to the $xy$ plane, and the bulk material
is  $\text{VO}_2$ with relative permittivity $\eps_{\text{VO}_2}$.
Provided that  wavelengths are much greater than the cross-sectional dimensions of the CTF columns, 
 the CTF may be regarded as a
homogeneous orthorhombic material which is characterized by the
frequency-domain constitutive relation  \c{HW,STF_Book,Vepa}
\begin{equation}
 \#D  =  \epso\,\=\eps_{\,\rm{CTF}}\. \#E\,. \label{epsbasic}
\end{equation}
Herein,  the CTF's relative permittivity dyadic is
\begin{equation}
\=\eps_{\,\rm{CTF}}= \=S_{\rm{y}} (\chi) \. \le \eps_{a} \,\uz\uz
+\eps_{b}\,\ux\ux \, +\,\eps_{c}\,\uy\uy \ri\.\=S_{\rm{y}}^{-1}(\chi)\,,
\label{ortho}
\end{equation}
 which has eigenvalues $\lec \eps_a, \eps_b, \eps_c \ric$, and $\epso$ is the permittivity of free space. The parallel columns of the CTF are inclined to the $xy$ plane at an  angle  $\chi \in(0,\pi/2]$, as prescribed by the
 inclination
dyadic
\begin{equation}
\=S_{\rm{y}}(\chi) = \uy\uy + (\ux\ux + \uz\uz) \, \cos\chi +
(\uz\ux-\ux\uz)\, \sin\chi\,.
\end{equation}
Note that $\eps_{a,b,c}$ and $\chi$ depend on the direction of the collimated vapor flux with respect to the
$xy$ plane \cite{HW}, and that $\eps_{a,b,c}$ also depend on $\lambdao$.

\begin{figure}[!htb]
\centering
\includegraphics[width=10cm]{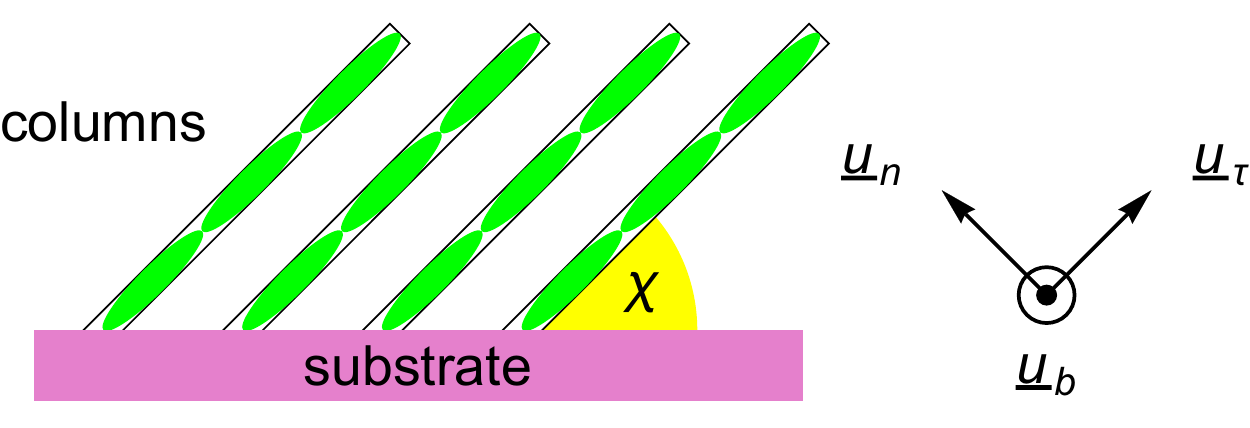}  
 \caption{\label{Fig1} Schematic representation of a CTF with columns inclined at angle $\chi$ to the planar substrate.
   }
\end{figure}

From the perspective of homogenization theory, each column of the CTF can be viewed as a string of elongated
ellipsoidal inclusions \cite{STF_Book,ML_JNP_2010}. All inclusions have the same orientation, and the same shape as
 prescribed by the shape dyadic
\begin{equation}
 \un \, \un + \gamma_\tau \, \ut \, \ut + \gamma_{\rm{b}} \, \ub \, \ub,
\end{equation}
wherein
\begin{equation}
\left. \begin{array}{l}
 \un = - \ux \, \sin \chi + \uz \, \cos \chi \vspace{4pt} \\
 \ut =  \ux \, \cos \chi + \uz \, \sin \chi \vspace{4pt} \\
\ub = - \uy
\end{array}
\right\}
\end{equation} are
the normal, tangential, and binormal basis vectors, respectively.
The columns of the CTF are needlelike.
Accordingly, the
shape parameters $\gamma_{\rm{b}} \gtrsim 1$ and $\gamma_\tau \gg 1$. For slim inclusions,
increasing $\gamma_\tau$ beyond 10 has negligible effect; hence,  $\gamma_\tau = 15$ is assumed.
 A schematic representation of such a CTF is presented in Fig.~\ref{Fig1}.

The void region between columns of the CTF is filled with air, which is taken to be electromagnetically equivalent to free space (or vacuum).
The fraction of the CTF's volume that is occupied by  ellipsoidal inclusions is denoted $f$. Since
  $f \in \le 0,1 \ri$, the porosity of the CTF is given as $1-f$.

\subsection{Bruggeman homogenization formalism}

The 
 eigenvalues $\lec \eps_a,\eps_b,\eps_c\ric$ of $\=\eps_{\,CTF}$ that characterize the CTF at macroscopic length scales
can be estimated from the parameters $\eps_{\text{VO}_2}$, $f$, and $\gamma_{\rm b}$ 
that characterize the CTF at microscopic length scales by means of the Bruggeman homogenization formalism.
The Bruggeman formalism 
has the advantage over other commonly used homogenization formalisms, e.g., the Maxwell Garnett formalism \c{Ward}, insofar as
it treats  the deposited
material region and the void region symmetrically. Consequently, there are no restrictions on the value of $f \in \le 0, 1 \ri$.

The polarizability density dyadic relevant to
  the deposited material (i.e., $\eps_{\text{VO}_2}$) is \c{MAEH}
\begin{eqnarray}
\nonumber \=a_{d} &=&  \les \eps_{\text{VO}_2} \, \=I - \le\eps_{a}
\,\uz\uz +\eps_{b}\,\ux\ux
\,  +\,\eps_{c}\,\uy\uy\ri\ris \\
&&\. \lec \=I +   \,\=D_{d} \. \les \eps_{\text{VO}_2}\, \=I -
\le\eps_{a} \,\uz\uz +\eps_{b}\,\ux\ux \, +\,\eps_{c}\,\uy\uy\ri\ris
\ric^{-1}\,,
\end{eqnarray}
while the polarizability density dyadic relevant to
  the void  medium (i.e., air) is
\begin{eqnarray}
\nonumber \=a_{v} &=&  \les  \=I - \le\eps_{a} \,\uz\uz
+\eps_{b}\,\ux\ux
\,  +\,\eps_{c}\,\uy\uy\ri\ris \\
&&\. \lec \=I + \=D_{v} \. \les  \=I - \le\eps_{a}
\,\uz\uz +\eps_{b}\,\ux\ux \, +\,\eps_{c}\,\uy\uy\ri\ris
\ric^{-1}\,,
\end{eqnarray}
wherein the corresponding depolarization dyadics
\begin{equation}
 \=D_{d} =
\,\frac{2}{\pi}\, \int_{\phi=0}^{\pi/2} \, d\phi
\int_{\theta=0}^{\pi/2}\, d\theta
\,  \sin \theta \, \, \frac{ \frac{ \cos^{2}
\theta}{\gamma_{\tau}^{2}}
  \,\, \ux \ux+\sin^{2}
\theta\left( \cos^{2} \phi \,\, \uz\uz +  \frac{ \sin^{2}
\phi}{\gamma_{b}^{2}}  \,\, \uy\uy \right) }{\epsilon_{b}\,\,
\frac{\cos^{2} \theta}{\gamma_{\tau}^{2}} +
   \sin^{2} \theta \left(\epsilon_{a} \cos^{2} \phi +
\epsilon_{c} \, \frac{\sin^{2} \phi}{\gamma_{b}^{2}} \right)} \,
\end{equation}
and
\begin{equation}
\=D_{v} =
\,\frac{2}{\pi}\, \int_{\phi=0}^{\pi/2} \, d\phi
\int_{\theta=0}^{\pi/2}\, d\theta
\,  \sin \theta \, \, \frac{ { \cos^{2} \theta}{}
  \,\, \ux \ux+\sin^{2}
\theta\left( \cos^{2} \phi \,\, \uz\uz +  { \sin^{2} \phi}{}  \,\,
\uy\uy \right) }{\epsilon_{b}\,\, {\cos^{2} \theta}{} +
   \sin^{2} \theta \left(\epsilon_{a} \cos^{2} \phi +
\epsilon_{c} \, {\sin^{2} \phi}{} \right)}.
\end{equation}
The essence of the Bruggeman formalism is that
 the volume-fraction-weighted  sum of the two polarizability density dyadics
is null valued \c{MAEH}; i.e.,
\begin{equation}
\label{Br}  f \,\=a_{d}+(1-f) \,\, \=a_{v} = \=0 \,,
\end{equation} 
with $\=0$ being the 3$\times$3 null dyadic.
The eigenvalues $\lec \eps_a,\eps_b,\eps_c\ric$ of $\=\eps_{\,CTF}$
can be extracted
from the Bruggeman eqn.~\r{Br}
 by numerical means, such as the Jacobi method \c{Michel_chap}, provided that  $  \eps_{\text{VO}_2}$, $ f$, 
 and $\gamma_{\rm{b}}$ are known \c{ML_JNP_2010}.

\section{Numerical studies}

\begin{figure}[!htb]
\centering
\includegraphics[width=8cm]{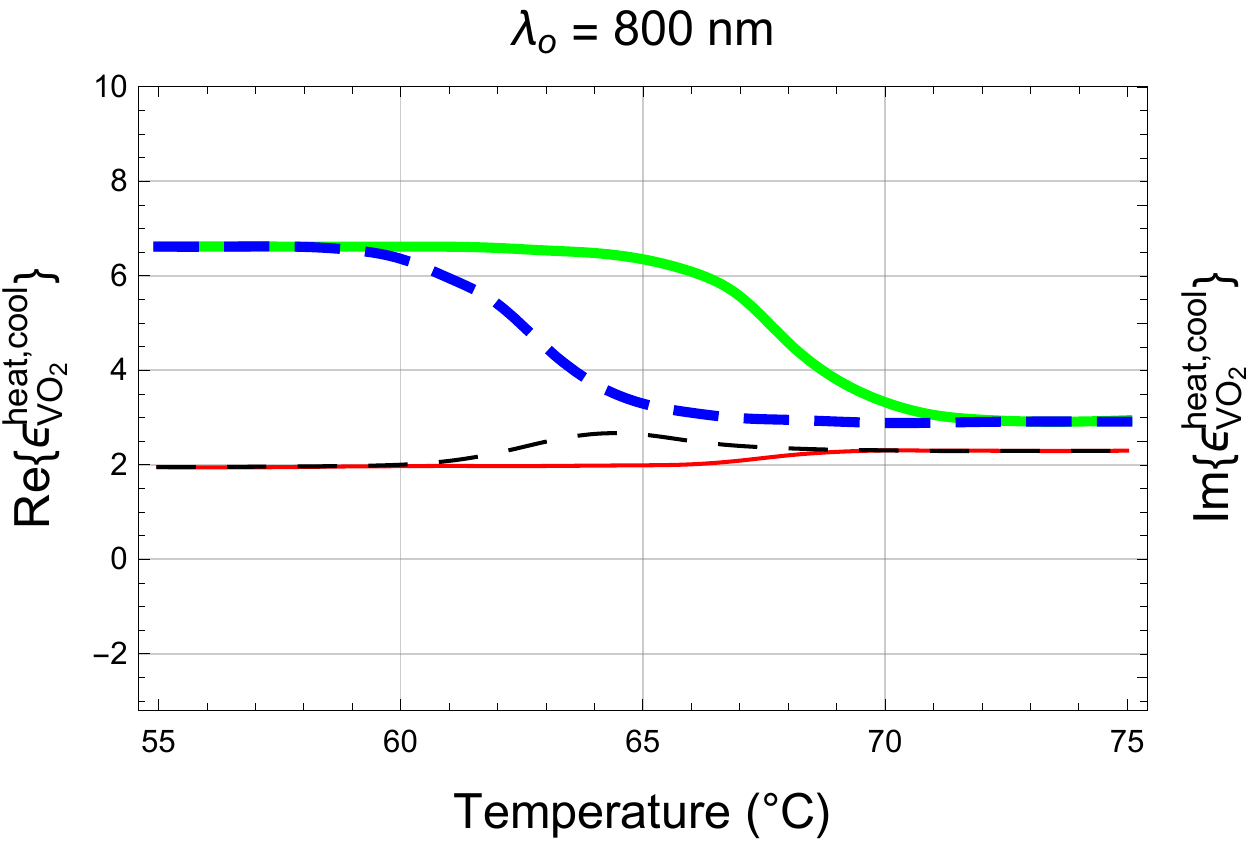} \hfill 
\includegraphics[width=8cm]{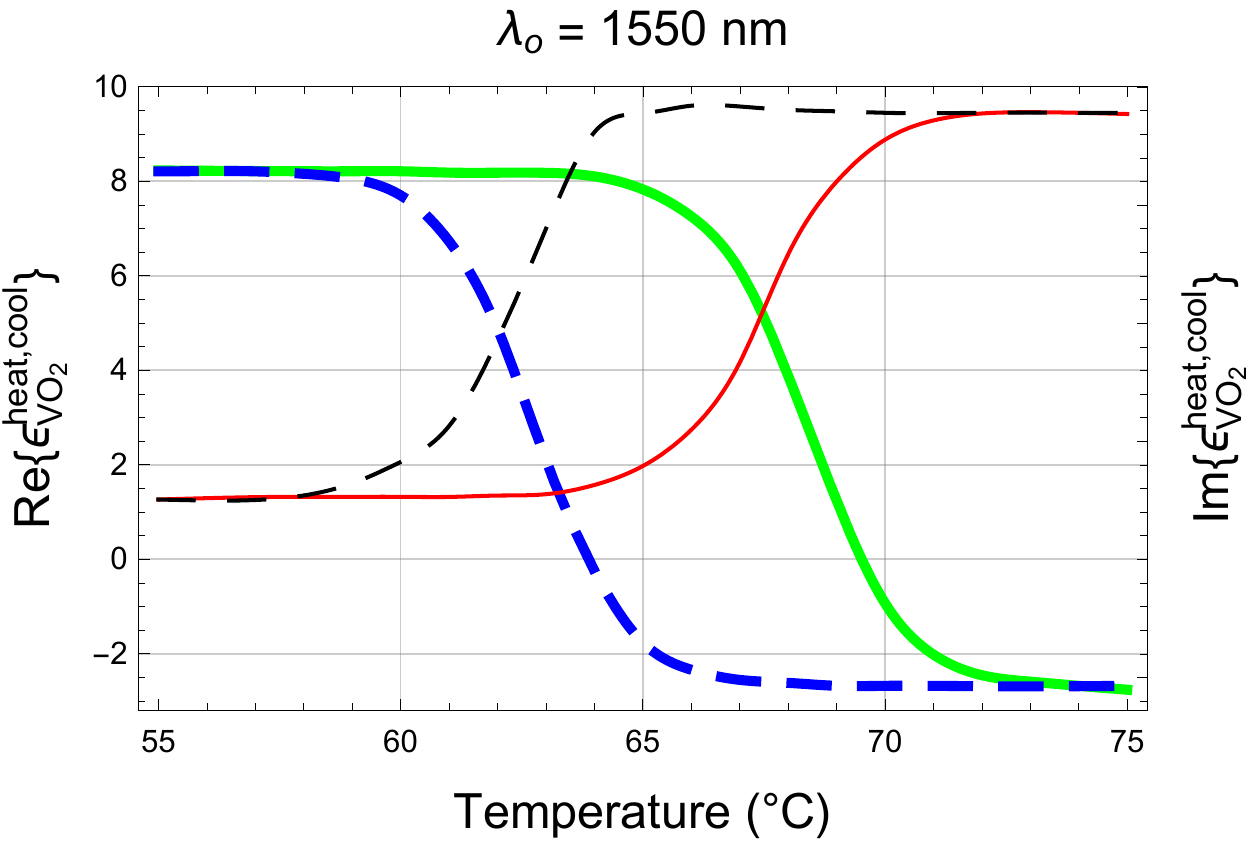}      \vspace{10mm}  \\
\includegraphics[width=8cm]{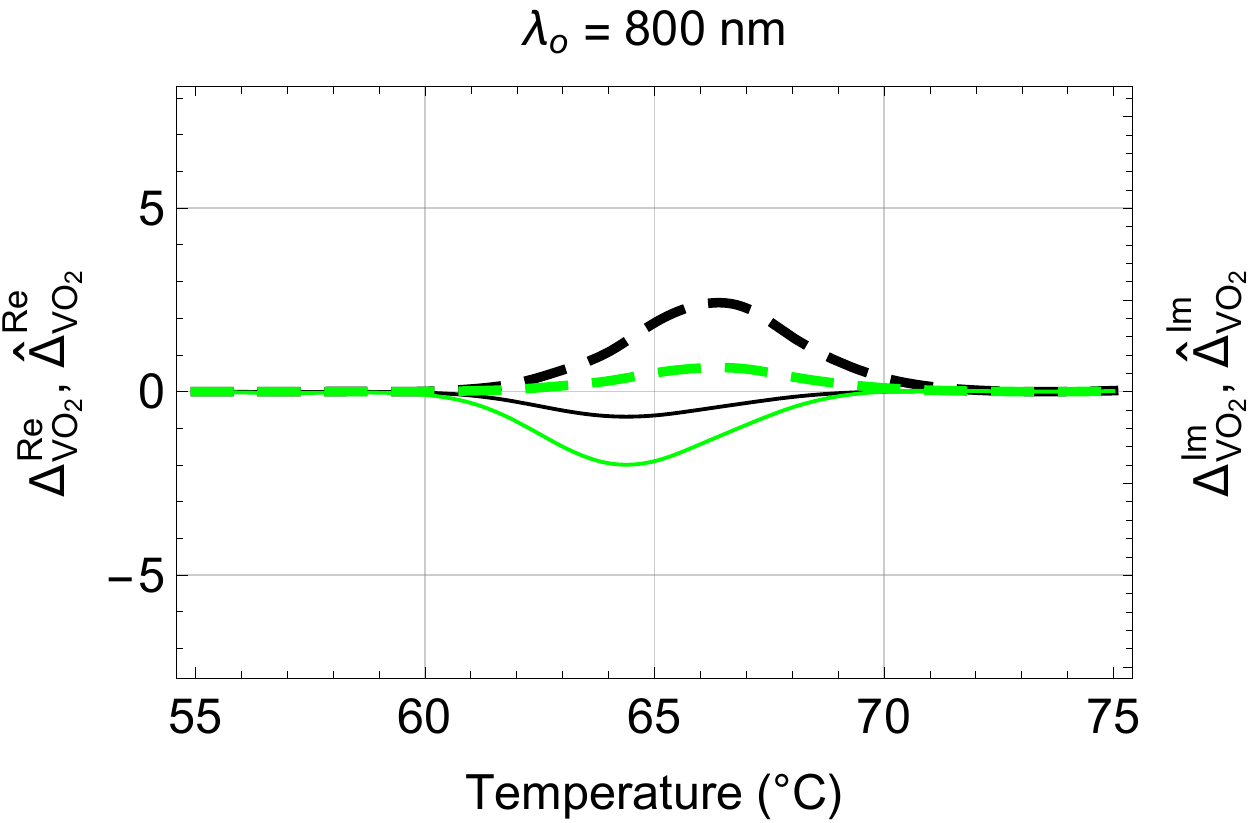}    \hfill  
\includegraphics[width=8cm]{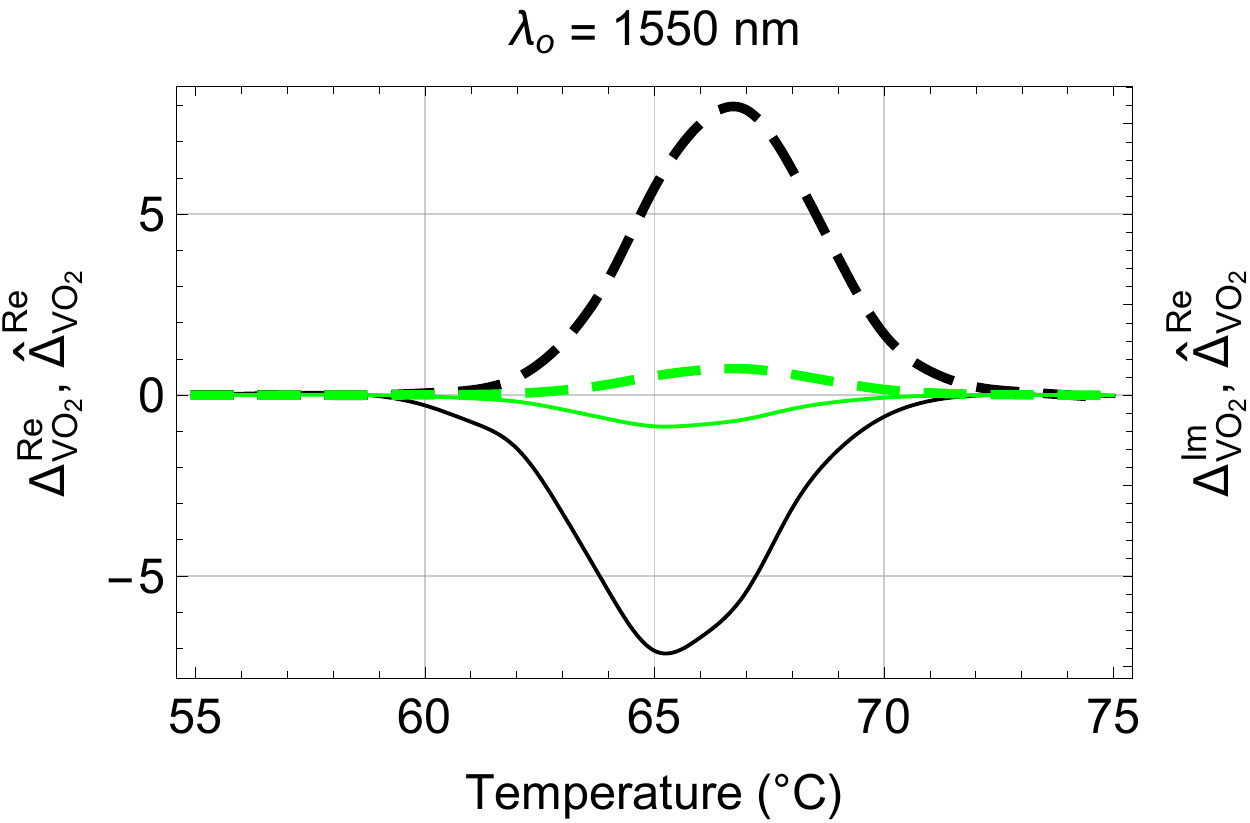} 
 \caption{\label{Fig2} Real and imaginary parts of $\eps^{\rm{heat,cool}}_{\text{VO}_2}$ plotted against temperature (derived from Fig.~4 of Ref.~\citenum{Cormier}), for $\lambdao \in \lec  800, 1550 \ric$ nm. 
 Also plotted against temperature are $\Delta^{\text{Re}}_{\text{VO}_2} $, $\hat{\Delta}^{\text{Re}}_{\text{VO}_2}$, $\Delta^{\text{Im}}_{\text{VO}_2}$, and $\hat{\Delta}^{\text{Im}}_{\text{VO}_2} $.
 Key: $\mbox{Re} \lec \eps^{\rm{heat}}_{\text{VO}_2} \ric$: thick green solid curve; $\mbox{Re} \lec \eps^{\rm{cool}}_{\text{VO}_2} \ric$: thick blue dashed curve; 
 $\mbox{Im} \lec \eps^{\rm{heat}}_{\text{VO}_2} \ric$: thin red solid curve; 
 $\mbox{Im} \lec \eps^{\rm{cool}}_{\text{VO}_2} \ric$: thin black dashed curve;  $\Delta^{\text{Re}}_{\text{VO}_2} $: thick black dashed curve;
 $\hat{\Delta}^{\text{Re}}_{\text{VO}_2} $: thick green dashed curve;
  $\Delta^{\text{Im}}_{\text{VO}_2} $
: thin black solid curve; $\hat{\Delta}^{\text{Im}}_{\text{VO}_2} $: thin green solid curve;   }
\end{figure}

The refractive index of $\text{VO}_2$ was   determined by Cormier \emph{et al.} \c{Cormier} for heating and cooling phases over the temperature range $55\degC$--$75\degC$, for   $\lambdao \in\left\{ 800,1550\right\}$~nm. 
Plots of the corresponding  real and imaginary parts of
$\eps_{\text{VO}_2}$ 
versus temperature are provided in Fig.~\ref{Fig2}, with $\eps^{\rm{heat}}_{\text{VO}_2}$ 
and  $\eps^{\rm{cool}}_{\text{VO}_2}$
denoting the values of $\eps_{\text{VO}_2}$ for the heating and cooling phases, respectively. 
Let $\eps^{55\degC}_{\text{VO}_2}$ and $\eps^{75\degC}_{\text{VO}_2}$ denote the values of $\eps_{\text{VO}_2}$ at $55\degC$ and $75\degC$, respectively.
Plots of the measures of hysteresis 
\begin{equation}
\left.
\begin{array}{l}
 \Delta^{\text{Re}}_{\text{VO}_2} = \mbox{Re} \lec \eps^{\rm{heat}}_{\text{VO}_2} \ric  - \mbox{Re} \lec \eps^{\rm{cool}}_{\text{VO}_2} \ric
 \vspace{8pt} \\
 \Delta^{\text{Im}}_{\text{VO}_2} = 
\mbox{Im} \lec \eps^{\rm{heat}}_{\text{VO}_2} \ric  - \mbox{Im} \lec \eps^{\rm{cool}}_{\text{VO}_2} \ric
\end{array}
\right\}
\end{equation}
and relative measures of hysteresis
\begin{equation}
\left.
\begin{array}{l}
\hat{ \Delta}^{\text{Re}}_{\text{VO}_2} = \displaystyle{
\frac{ \Delta^{\text{Re}}_{\text{VO}_2} }
{ \left|
\mbox{Re} \lec \eps^{75\degC}_{\text{VO}_2} \ric - \mbox{Re} \lec \eps^{55\degC}_{\text{VO}_2} \ric \right|}} \vspace{8pt} \\
\hat{
 \Delta}^{\text{Im}}_{\text{VO}_2} = \displaystyle{
\frac{ \Delta^{\text{Im}}_{\text{VO}_2} }
{ \left|
\mbox{Im} \lec \eps^{75\degC}_{\text{VO}_2} \ric - \mbox{Im} \lec \eps^{55\degC}_{\text{VO}_2} \ric \right|}} 
\end{array}
\right\}
\end{equation}
versus temperature are also provided in Fig.~\ref{Fig2}.

For $\lambdao =$ 800 nm,  $\mbox{Re} \lec \eps_{\text{VO}_2} \ric$ 
and  $\mbox{Im} \lec \eps_{\text{VO}_2} \ric$
exhibit hysteresis 
over the approximate temperature range $60\degC$--$71\degC$.
In the  thermal-hysteresis regime, $\mbox{Re} \lec \eps^{\rm{heat}}_{\text{VO}_2} \ric >
\mbox{Re} \lec \eps^{\rm{cool}}_{\text{VO}_2} \ric$ whereas
 $\mbox{Im} \lec \eps^{\rm{heat}}_{\text{VO}_2} \ric <
\mbox{Im} \lec \eps^{\rm{cool}}_{\text{VO}_2} \ric$. Also, 
the maximum magnitude of  
 $\Delta^{\text{Re}}_{\text{VO}_2}$
 is substantially larger than the maximum magnitude of 
 $ \Delta^{\text{Im}}_{\text{VO}_2}$
but
the maximum magnitude of  
 $\hat{\Delta}^{\text{Re}}_{\text{VO}_2}$
 is substantially smaller than the maximum magnitude of 
 $ \hat{\Delta}^{\text{Im}}_{\text{VO}_2}$. That is, 
 the hysteresis is more pronounced for the real part of 
 $ \eps_{\text{VO}_2}$ than for the imaginary part of $ \eps_{\text{VO}_2}$,
but
 the relative hysteresis is more pronounced for the imaginary part of 
 $ \eps_{\text{VO}_2}$ than for the real part of $ \eps_{\text{VO}_2}$.

The nature of the thermal hysteresis for  $\lambdao =$ 1550 nm is qualitatively similar to that for $\lambdao =$ 800 nm, but
\begin{itemize}
\item
the thermal-hysteresis regime   is
 slightly wider;
 \item
 the maximum magnitudes of  $ \Delta^{\text{Re}}_{\text{VO}_2} $ and $ \Delta^{\text{Im}}_{\text{VO}_2} $ are substantially larger;
 \item
 the maximum magnitude of  $ \hat{\Delta}^{\text{Re}}_{\text{VO}_2} $ is slightly larger and the maximum magnitude of
 $\hat{\Delta}^{\text{Im}}_{\text{VO}_2}$ is substantially smaller 
\end{itemize}
for  $\lambdao =$ 1550 nm than for  $\lambdao =$ 800 nm. Also,  for  $\lambdao =$ 1550 nm, the maximum magnitude of  $ \Delta^{\text{Re}}_{\text{VO}_2} $ is approximately the same as that of  $\Delta^{\text{Im}}_{\text{VO}_2}$
and likewise  the maximum magnitude of  $ \hat{\Delta}^{\text{Re}}_{\text{VO}_2} $ is approximately the same as that of  $\hat{\Delta}^{\text{Im}}_{\text{VO}_2}$; 
 i.e., the hysteresis and  relative hysteresis are approximately equally pronounced for the real and  imaginary parts of 
 $ \eps_{\text{VO}_2}$.
Another  notable difference is that 
$\mbox{Re} \lec \eps^{\rm{heat}}_{\text{VO}_2} \ric >0$ and $\mbox{Re} \lec \eps^{\rm{cool}}_{\text{VO}_2} \ric >0$
across the entire temperature range considered
for $\lambdao =$ 800 nm but, for  $\lambdao =$ 1550 nm, $\mbox{Re} \lec \eps^{\rm{heat}}_{\text{VO}_2} \ric >0$ in the temperature range $55\degC$--$69.5\degC$ and $\mbox{Re} \lec \eps^{\rm{heat}}_{\text{VO}_2} \ric <0$ at higher temperatures, and
$\mbox{Re} \lec \eps^{\rm{cool}}_{\text{VO}_2} \ric >0$ in the temperature range $55\degC$--$63.5\degC$ and $\mbox{Re} \lec \eps^{\rm{cool}}_{\text{VO}_2} \ric <0$ at higher temperatures.
That is,  VO${}_2$ is a dissipative dielectric material  across the entire temperature range considered for $\lambdao =$ 800 nm, but 
it is a dissipative dielectric material  at  low temperatures and becomes metal-like at higher temperatures for $\lambdao =$ 1550 nm.
Furthermore, the temperature at which the transition between as dissipative dielectric material and a metal-like material   occurs depends upon whether the material is being heated or cooled.

\begin{figure}[!htb]
\centering
\includegraphics[width=8cm]{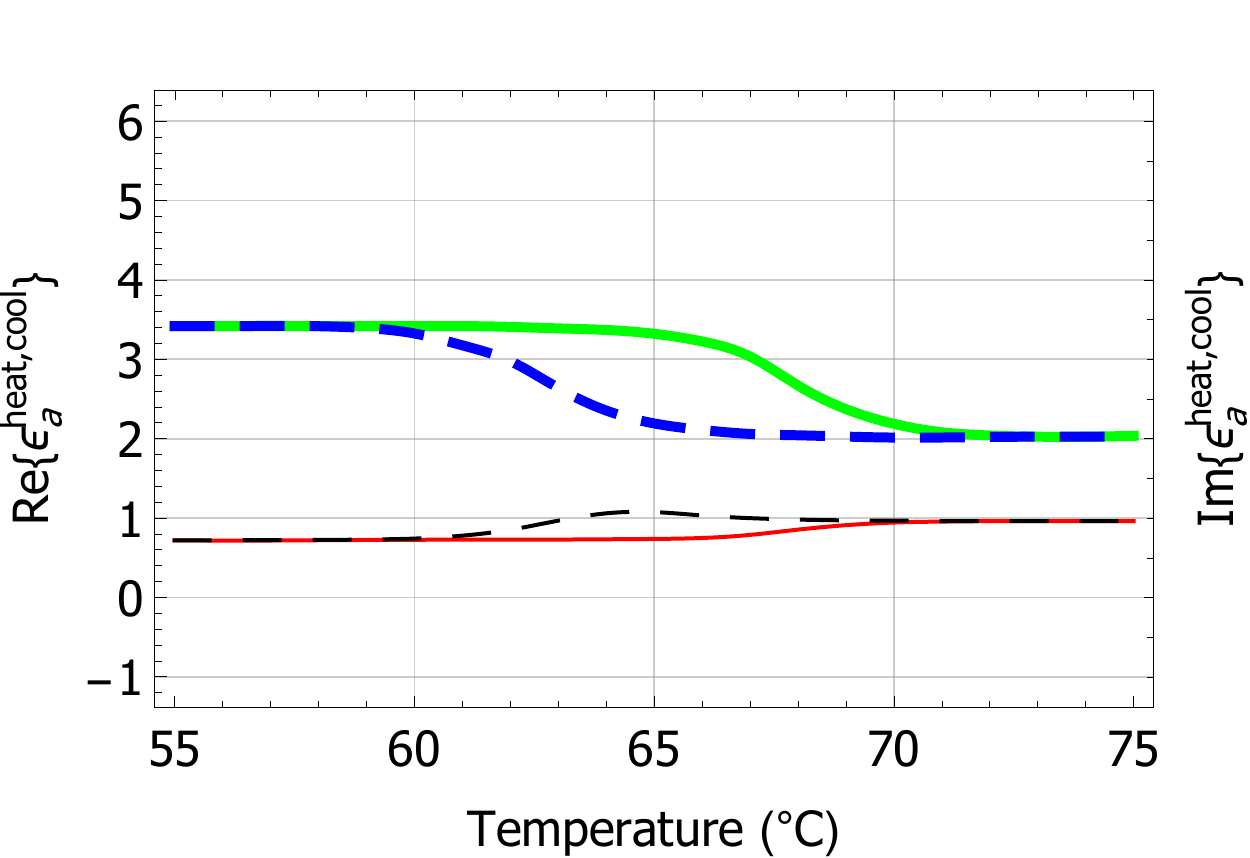}    
\hfill \includegraphics[width=8cm]{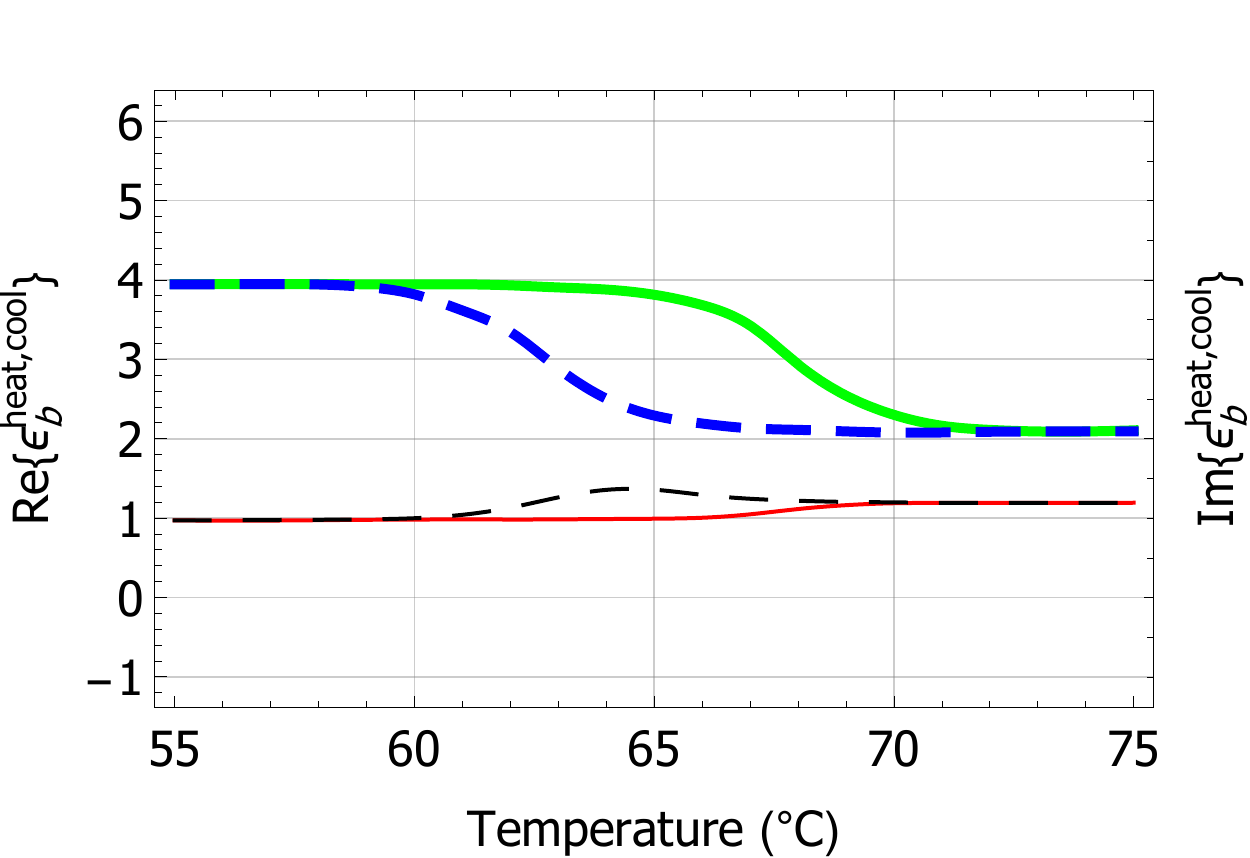}    
 \vspace{10mm}  \\
\includegraphics[width=8cm]{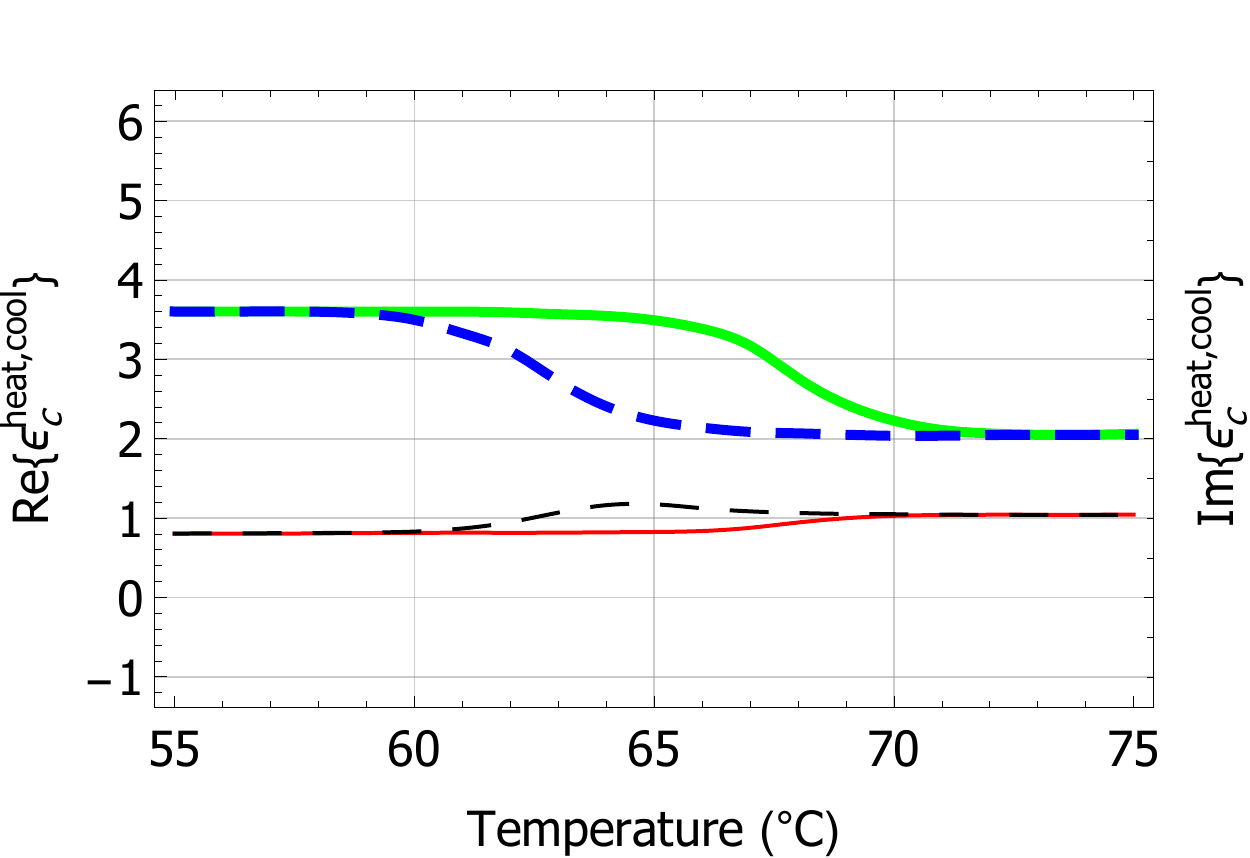}    
\hfill \includegraphics[width=8cm]{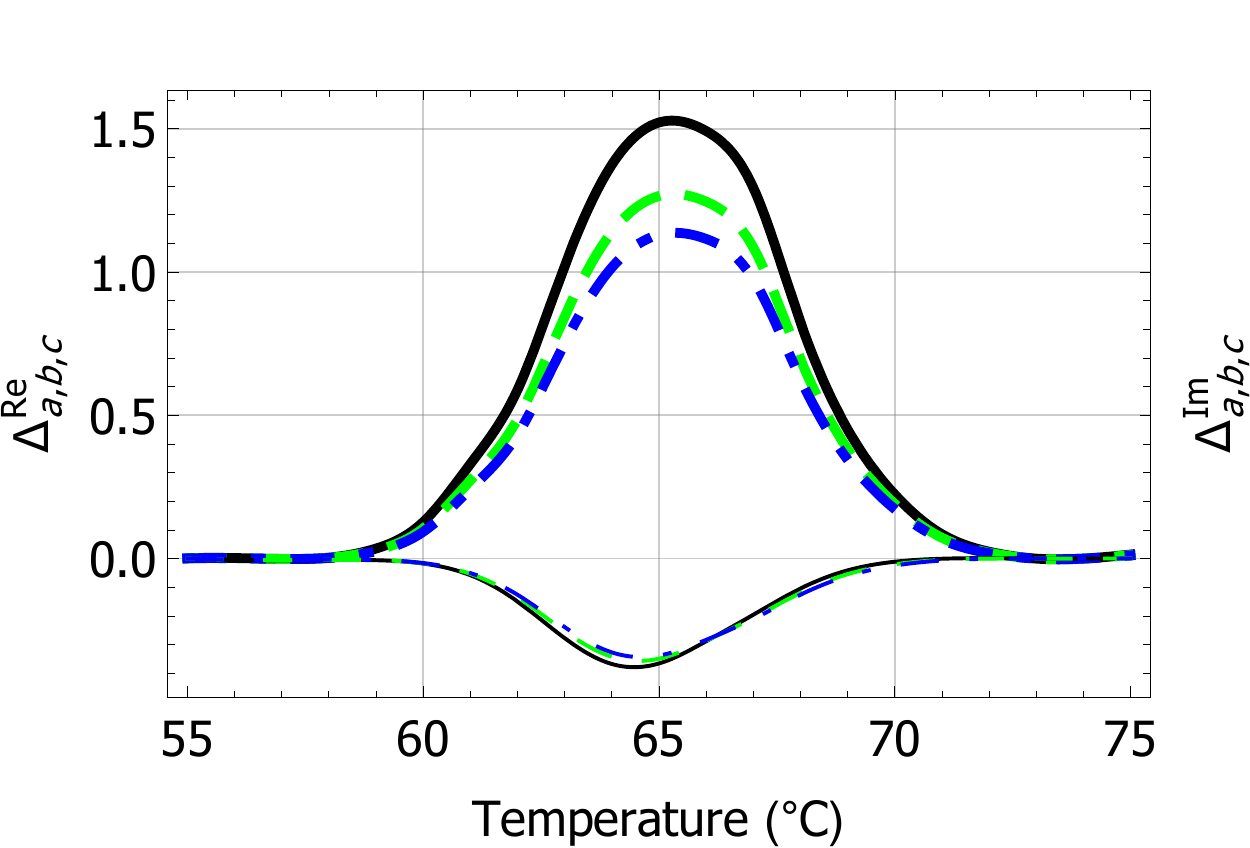}    \vspace{10mm}  \\
\includegraphics[width=8cm]{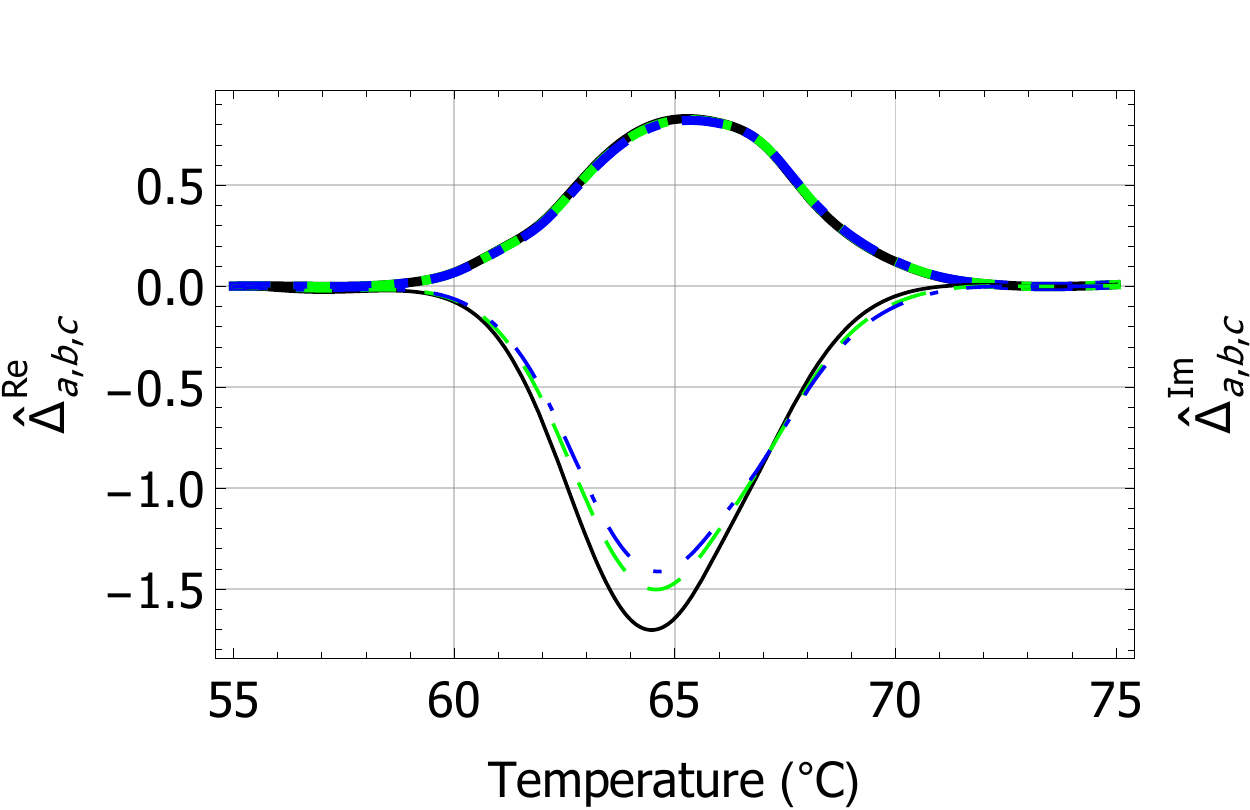}    
 \caption{\label{Fig3} Real and imaginary parts of the CTF's relative permittivity scalars $\eps^{\rm{heat,cool}}_{a,b,c}$  estimated using the
  Bruggeman homogenization formalism 
  plotted against temperature
  for $\lambdao= 800$ nm, with $\gamma_\tau = 15$, $\gamma_b = 1.5$, and $f = 0.6$.
  Also plotted against temperature are $\Delta^{\text{Re}}_{a,b,c} $, $\Delta^{\text{Im}}_{a,b,c} $, $\hat{\Delta}^{\text{Re}}_{a,b,c} $ and $\hat{\Delta}^{\text{Im}}_{a,b,c}$.
   Key:
 $\mbox{Re} \lec \eps^{\rm{heat}}_{a,b,c} \ric$: thick green solid curve; $\mbox{Re} \lec \eps^{\rm{cool}}_{a,b,c} \ric$: thick blue dashed curve; 
 $\mbox{Im} \lec \eps^{\rm{heat}}_{a,b,c} \ric$: thin red solid curve; 
 $\mbox{Im} \lec \eps^{\rm{cool}}_{a,b,c} \ric$: thin black dashed curve;  
   $ \Delta^{\text{Re}}_{a} $, $ \hat{\Delta}^{\text{Re}}_{a} $: thick  blue broken dashed curve;
   $ \Delta^{\text{Re}}_{b} $, $\hat{ \Delta}^{\text{Re}}_{b} $: thick black solid curve;
$ \Delta^{\text{Re}}_{c} $, $ \hat{\Delta}^{\text{Re}}_{c} $: thick green dashed curve;
      $ \Delta^{\text{Im}}_{a} $, $ \hat{\Delta}^{\text{Im}}_{a} $: thin  blue broken dashed curve;
   $ \Delta^{\text{Im}}_{b} $, $ \hat{\Delta}^{\text{Im}}_{b} $: thin black solid curve;
$ \Delta^{\text{Im}}_{c} $, $ \hat{\Delta}^{\text{Im}}_{c} $: thin green dashed curve.
   }
\end{figure}

Next we turn to the   VO${}_2$ CTFs. We begin by setting $\gamma_{\rm{b}}= 1.5$ and $f=0.6$, which are  typical values for CTFs made from a range of different materials \c{ML_JNP_2010}; the effects of varying  $\gamma_{\rm{b}}$ and $f$ are explored later in this section. For $\lambdao =$ 800 nm, plots of the Bruggeman estimates of the real and imaginary parts of 
$\eps_a$, $\eps_b$, and $\eps_c$ (for both heating and cooling phases) versus temperature are  presented in Fig.~\ref{Fig3}.
Therein,
$\eps^{\rm{heat}}_{a,b,c}$ 
and  $\eps^{\rm{cool}}_{a,b,c}$
denote the values of $\eps_{a,b,c}$ for the heating and cooling phases, respectively.

 The orthorhombic symmetry of the CTF is apparent  since, for both heating and cooling phases, the values of $\eps_b$ are clearly different  from the values of $\eps_a$ and $\eps_c$, and there are smaller differences between the values of $\eps_a$ and $\eps_c$. In addition, both the real and imaginary parts of $\eps_a$, $\eps_b$, and $\eps_c$ exhibit thermal hysteresis to varying degrees. Plots of the measures of hysteresis 
 \begin{equation}
\left.
\begin{array}{l}
 \Delta^{\text{Re}}_{\ell}  = \mbox{Re} \lec \eps^{\rm{heat}}_\ell \ric  - \mbox{Re} \lec \eps^{\rm{cool}}_\ell \ric
 \vspace{8pt} \\
 \Delta^{\text{Im}}_\ell = 
\mbox{Im} \lec \eps^{\rm{heat}}_\ell \ric  - \mbox{Im} \lec \eps^{\rm{cool}}_\ell \ric 
\end{array}
\right\}, \qquad  \ell \in \lec a, b, c \ric
\end{equation}
and relative  measures of hysteresis
\begin{equation}
\left.
\begin{array}{l}
 \hat{\Delta}^{\text{Re}}_{\ell}  = \displaystyle{
\frac{\Delta^{\text{Re}}_{\ell}}
{ \left|
\mbox{Re} \lec \eps^{75\degC}_\ell \ric - \mbox{Re} \lec \eps^{55\degC}_\ell \ric \right|}} \vspace{8pt} \\
 \hat{\Delta}^{\text{Im}}_\ell = \displaystyle{
\frac{\Delta^{\text{Im}}_{\ell}}
{ \left|
\mbox{Im} \lec \eps^{75\degC}_\ell \ric - \mbox{Im} \lec \eps^{55\degC}_\ell \ric \right|}} 
\end{array}
\right\}, \qquad  \ell \in \lec a, b, c \ric
\end{equation}
versus temperature are also presented in Fig.~\ref{Fig3}.
 Herein, $\eps^{55\degC}_\ell$ and $\eps^{75\degC}_\ell$ denote the values of $\eps_\ell$ at $55\degC$ and $75\degC$, respectively, with $\ell \in \lec a, b, c \ric$.
 The maximum magnitude of   $ \Delta^{\text{Re}}_b$ is substantially larger than the maximum magnitude of  $ \Delta^{\text{Re}}_c$ which, in turn, is substantially larger than the maximum magnitude of  $ \Delta^{\text{Re}}_a$.
Also,  the   maximum magnitude of  $\Delta^{\text{Im}}_b$ is slightly larger than the maximum magnitude of  
$ \Delta^{\text{Im}}_c$ which, in turn, is slightly larger than the maximum magnitude of  $\Delta^{\text{Im}}_a$.
As regards the relative measures of hysteresis,
the maximum of $| \hat{\Delta}^{\text{Im}}_b|$ is substantially larger than the maximum of  $| \hat{\Delta}^{\text{Im}}_c|$ which, in turn, is substantially larger than the maximum of  $| \hat{\Delta}^{\text{Re}}_a|$; in contrast, the plots for $ \hat{\Delta}^{\text{Re}}_a$ , $ \hat{\Delta}^{\text{Re}}_b$, and $ \hat{\Delta}^{\text{Re}}_c$ are all similar.
Thus,     thermal hysteresis   has been  transformed from isotropic for bulk  VO${}_2$  to anisotropic for the VO${}_2$  CTF.

\begin{figure}[!htb]
\centering
\includegraphics[width=8cm]{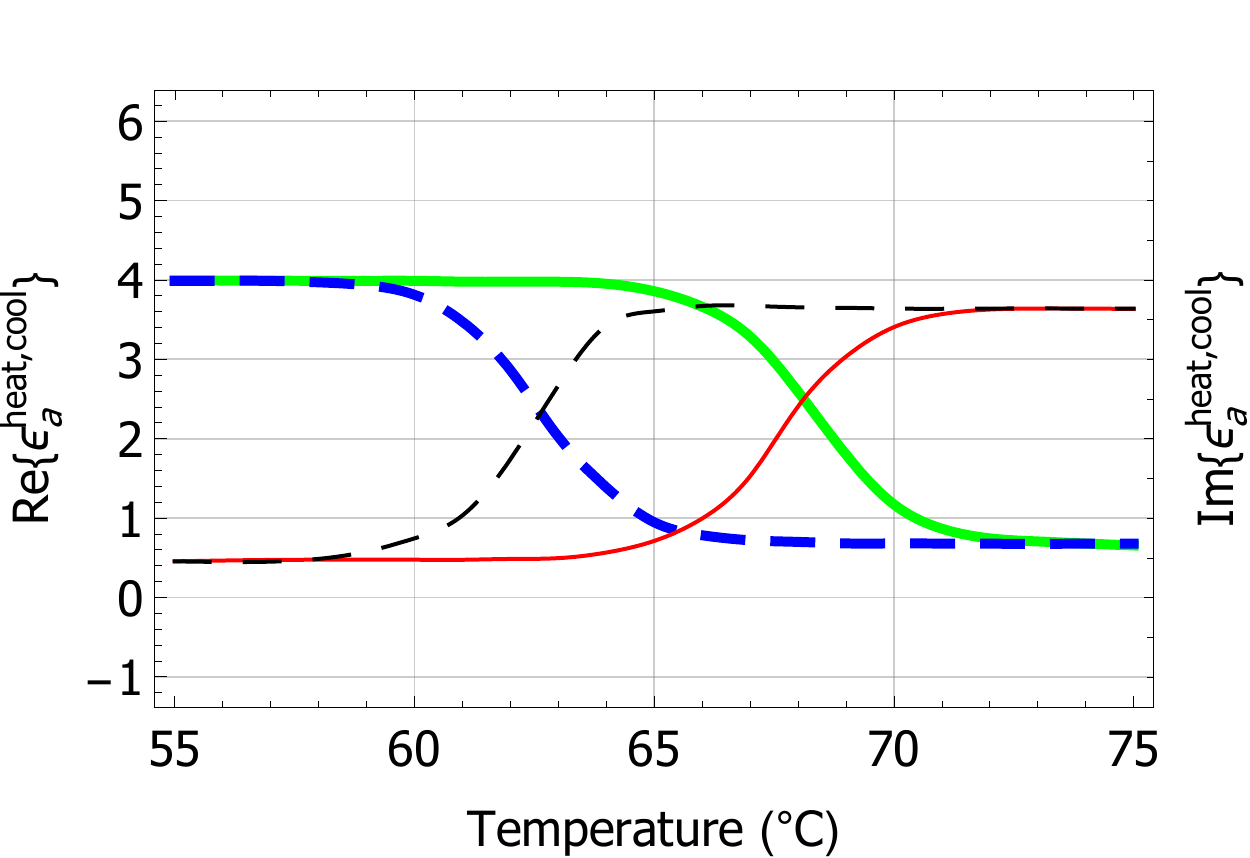}    
\hfill \includegraphics[width=8cm]{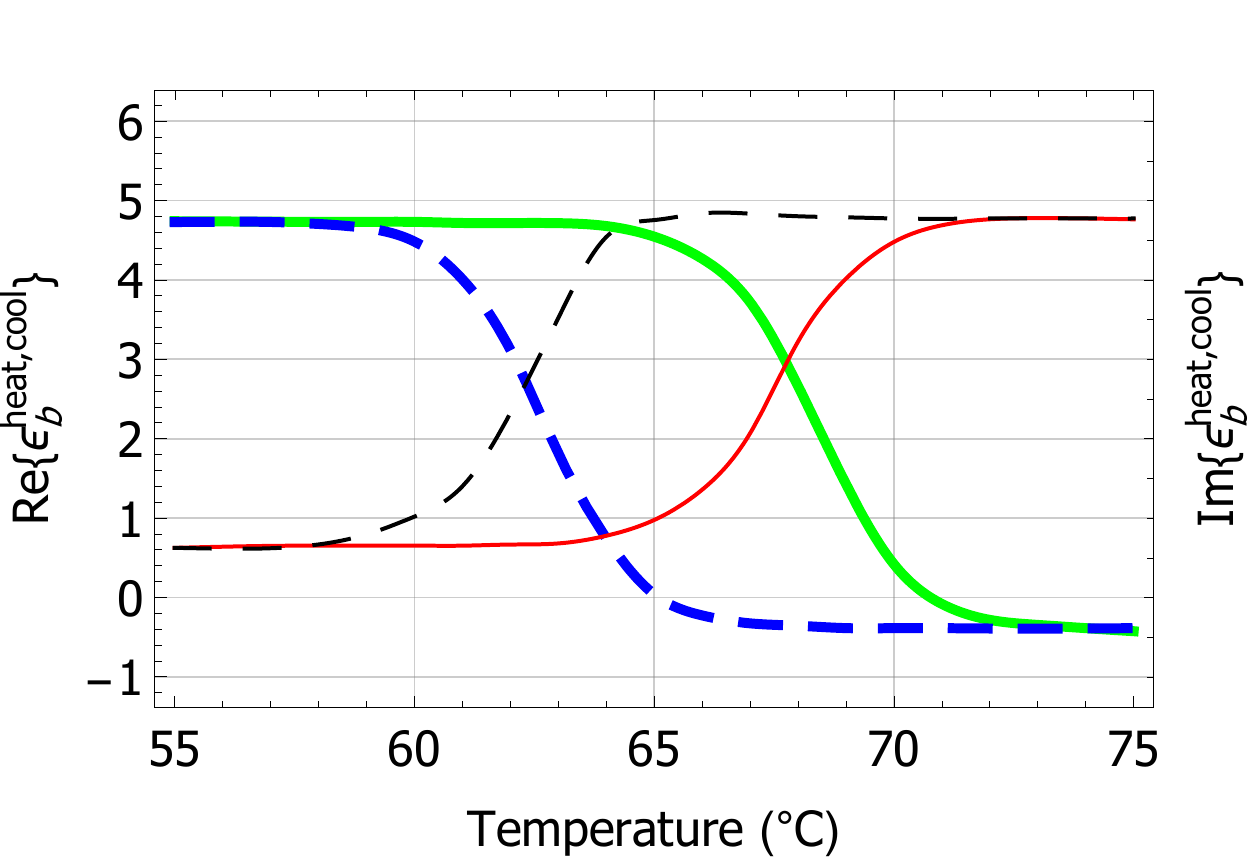}    
 \vspace{10mm}  \\
\includegraphics[width=8cm]{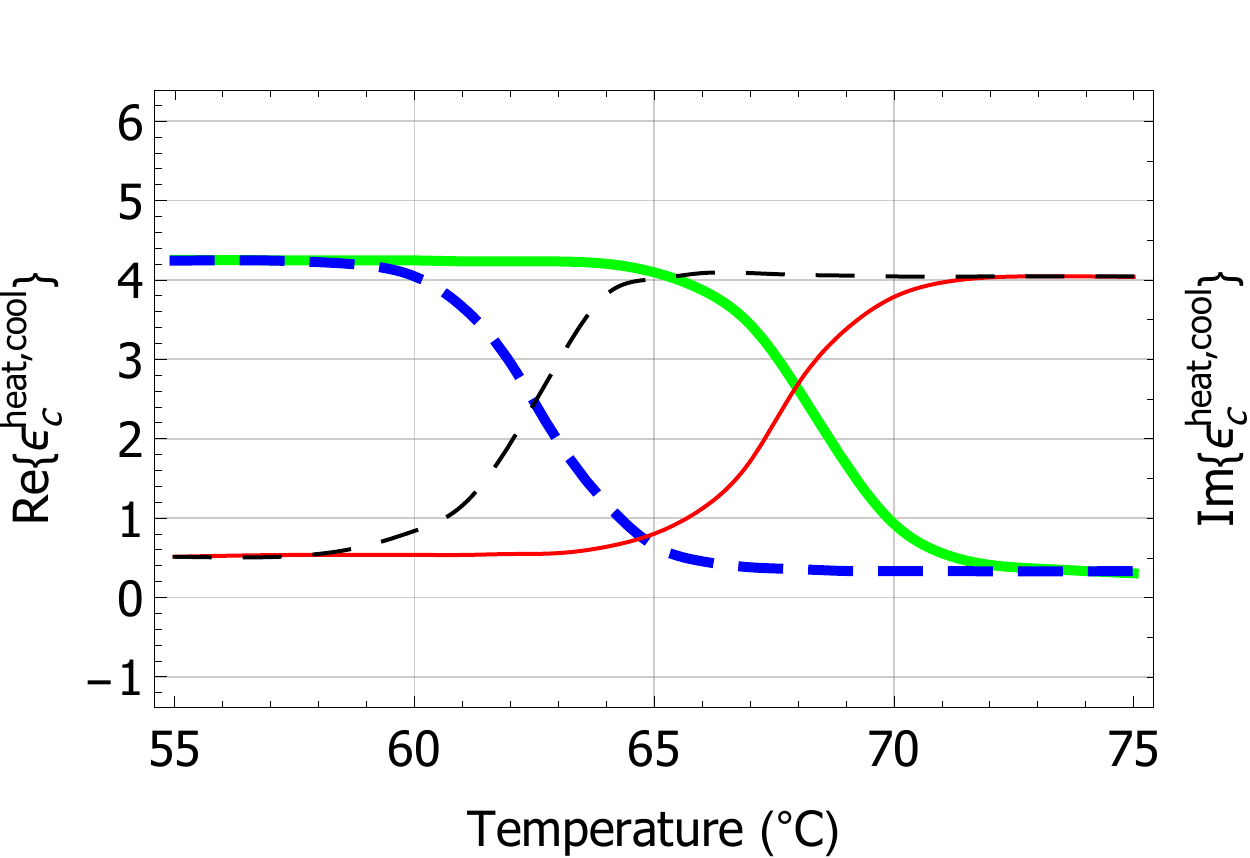}
\hfill \includegraphics[width=8cm]{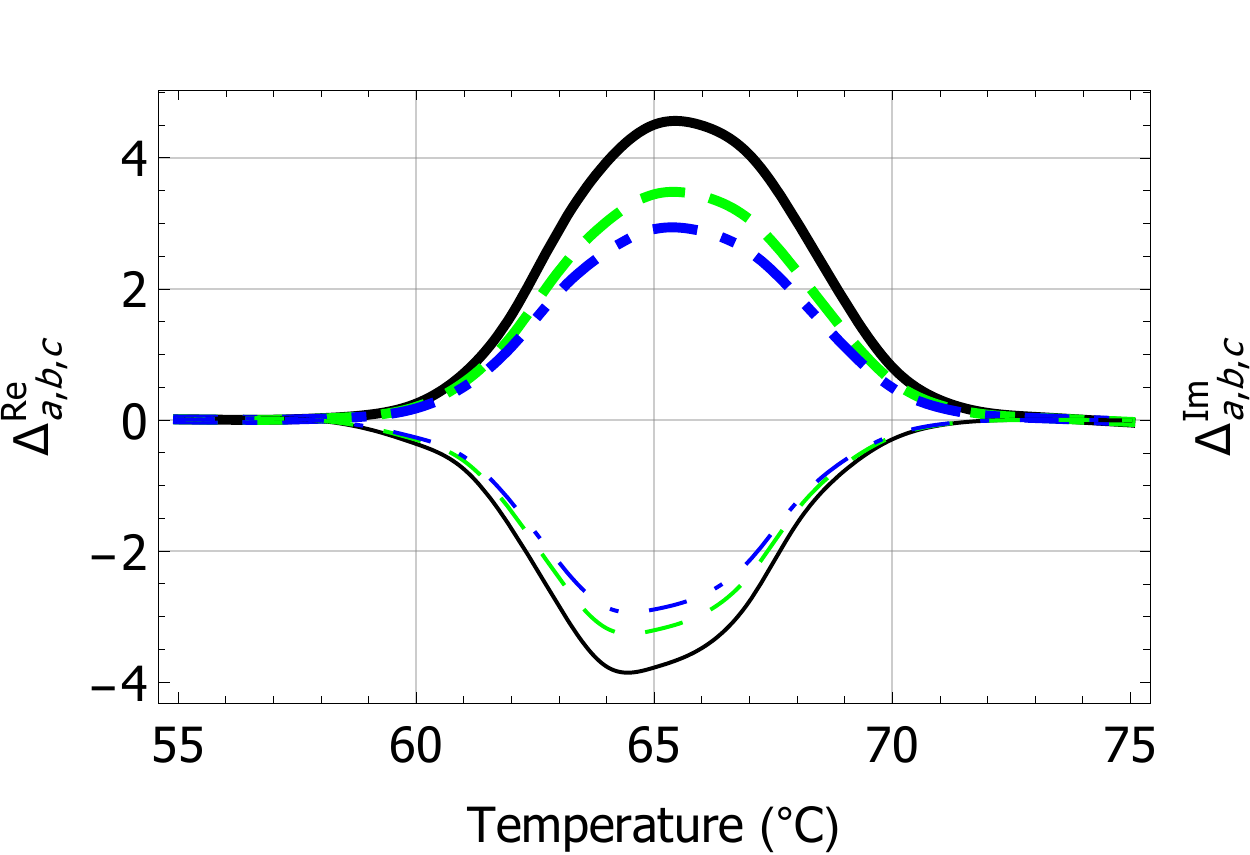}    
 \vspace{10mm}  \\
\includegraphics[width=8cm]{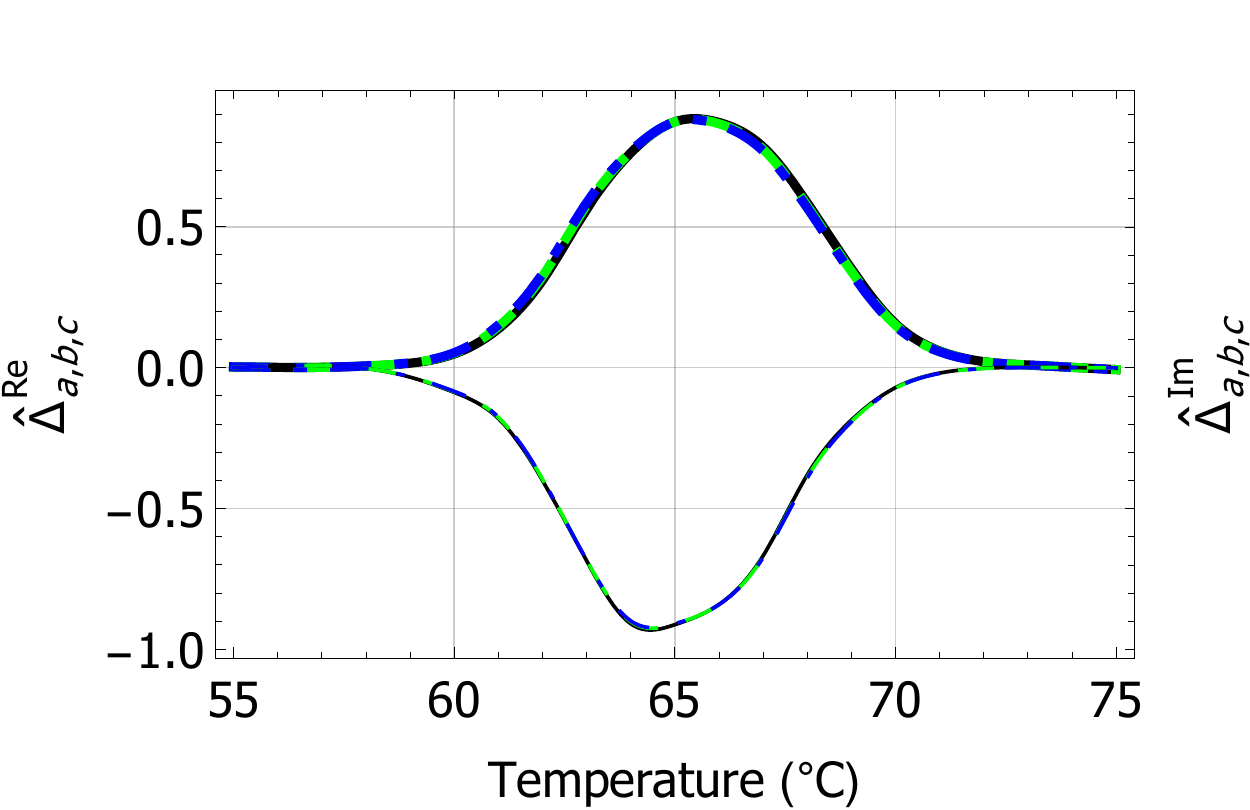}
     \caption{\label{Fig4} As Fig.~\ref{Fig3} but for $\lambdao= 1550$ nm.
   }
\end{figure}

The computations for Fig.~\ref{Fig3} were repeated for $\lambdao =$ 1550 nm and the results plotted in Fig.~\ref{Fig4}. As regards anisotropy and hysteresis, the results   for 
$\lambdao =$ 1550 nm displayed in Fig.~\ref{Fig4} are qualitatively similar to those for $\lambdao =$ 800 nm displayed in Fig.~\ref{Fig3}, but 
\begin{itemize}
\item
the anisotropy of the CTF is substantially more pronounced
and
\item  the anisotropy of the thermal hysteresis,   but not of the relative thermal hysteresis, is substantially more pronounced
\end{itemize}
 for $\lambdao =$ 1550 nm. 
 
 However, there is an important difference in the nature of the  VO${}_2$   CTF for
 $\lambdao =$ 1550 nm as compared to its nature for $\lambdao =$ 800 nm. According to Fig.~\ref{Fig3},
the real parts of $\eps_a$, $\eps_b$, and $\eps_c$ (for both heating and cooling phases) are all positive valued for $\lambdao =$ 800 nm across the entire temperature range considered. 
According to Fig.~\ref{Fig4},
the same is true for  $\eps_a$ and $\eps_c$  for $\lambdao =$ 1550 nm, but $\mbox{Re} \lec  \eps^{\rm{heat}}_b \ric $ is positive valued only for  the temperature range $55\degC$--$71\degC$ and negative valued at higher temperatures,  and $\mbox{Re} \lec  \eps^{\rm{cool}}_b \ric $ is positive valued only for  the temperature range $55\degC$--$65\degC$ and negative valued at higher temperatures.
That is, the CTF is a dissipative dielectric material for $\lambdao =$ 800 nm  across the entire temperature range considered,
it is a dissipative dielectric material only at  low temperatures and becomes a hyperbolic material \c{Smolyaninov} at higher temperatures for $\lambdao =$ 1550 nm.
Furthermore, the temperature at which the  dissipative dielectric material/hyperbolic material transition occurs depends upon whether the material is being heated or cooled.

\begin{figure}[!htb]
\centering
\includegraphics[width=4.9cm]{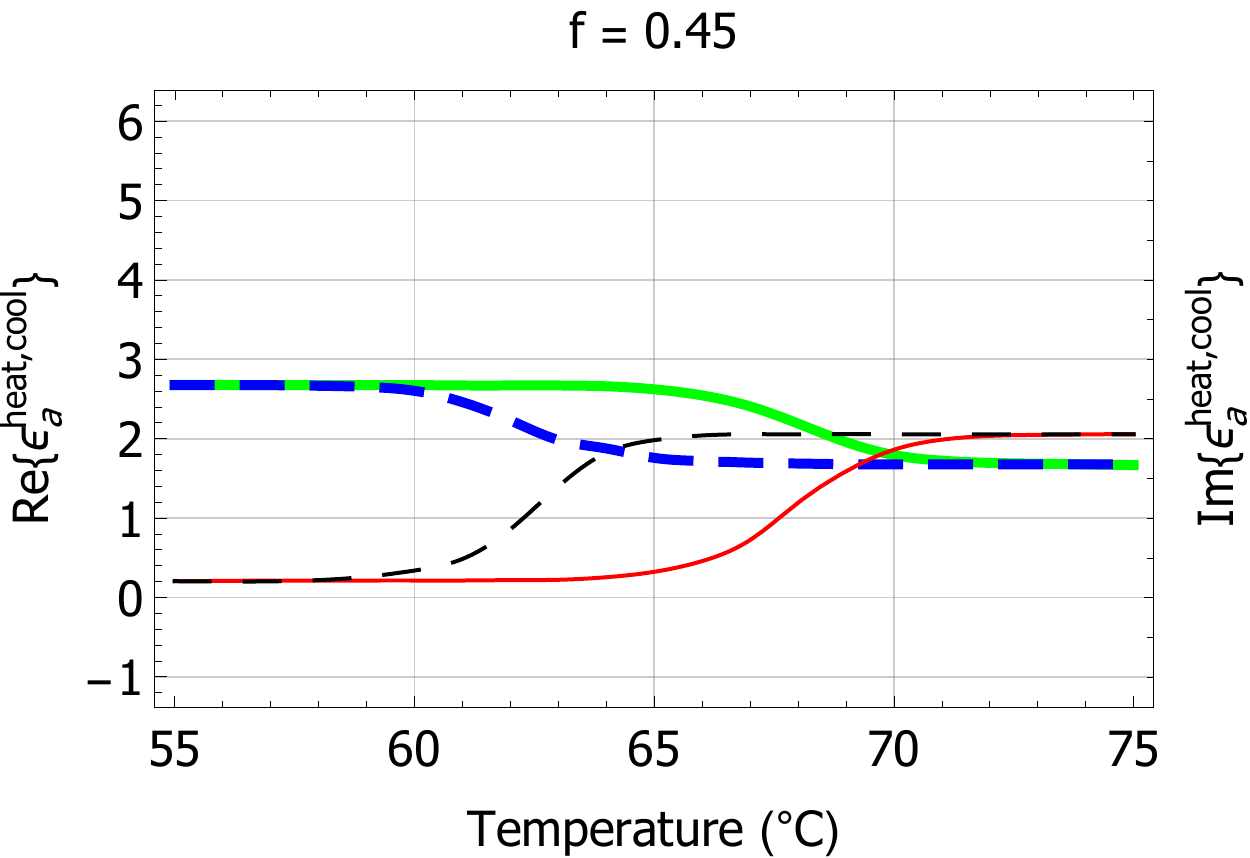}    \hspace{2mm}
\includegraphics[width=4.9cm]{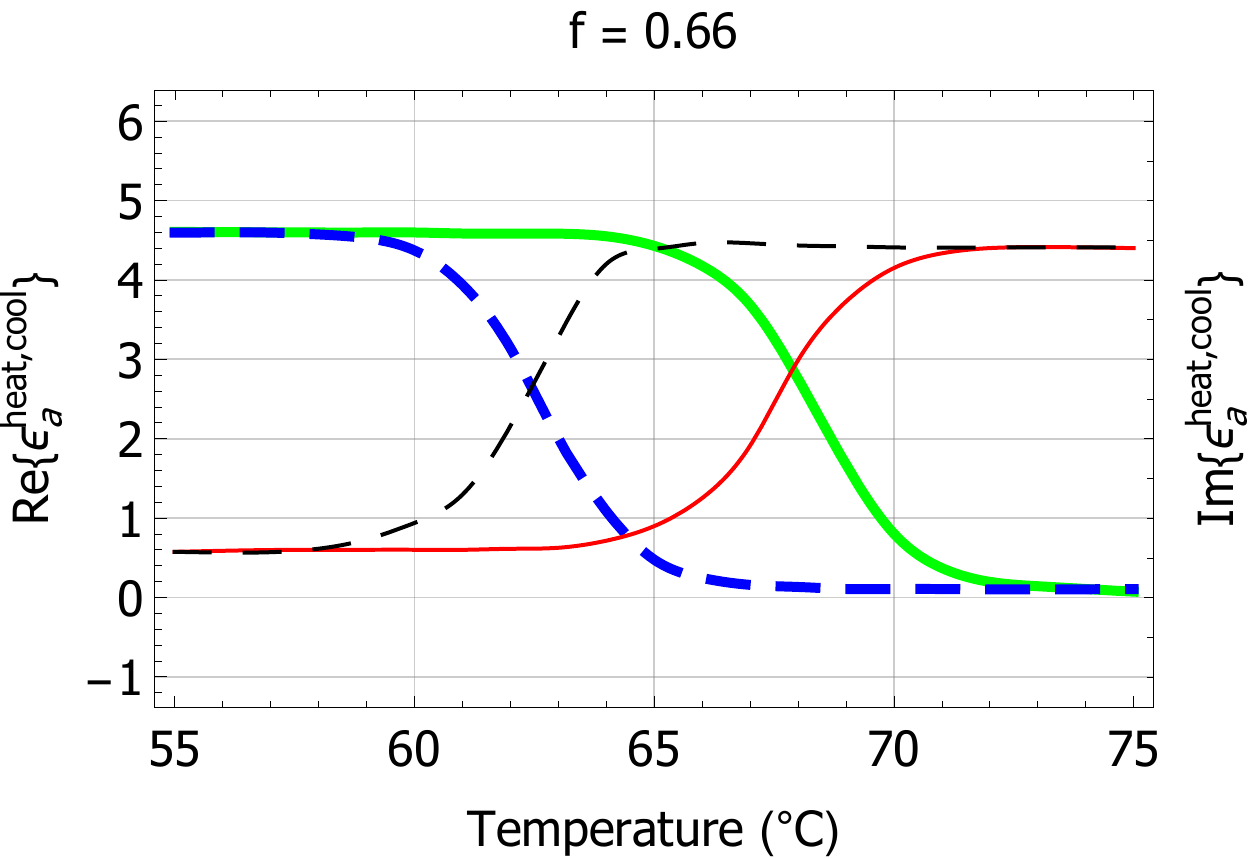}  \hspace{2mm}
\includegraphics[width=4.9cm]{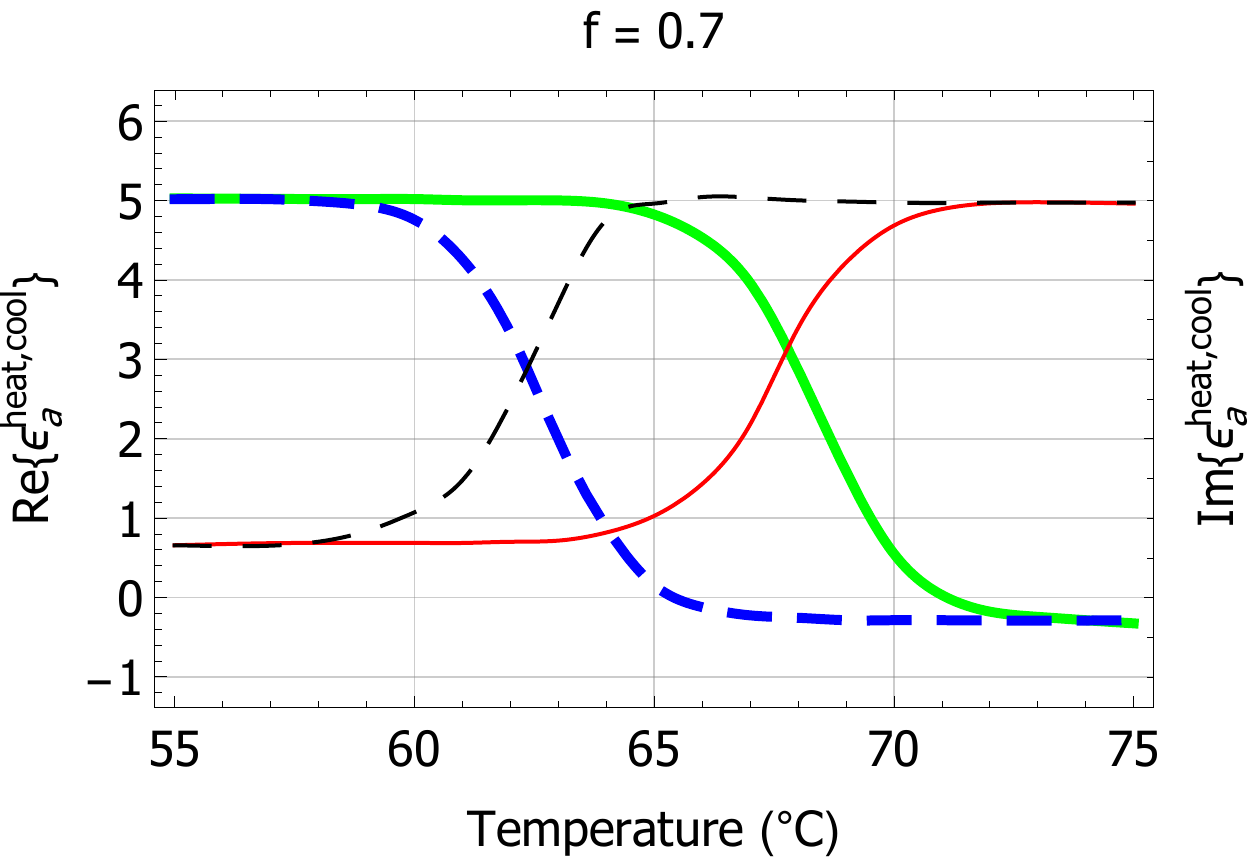}   
 \vspace{5mm}  \\
 \includegraphics[width=4.9cm]{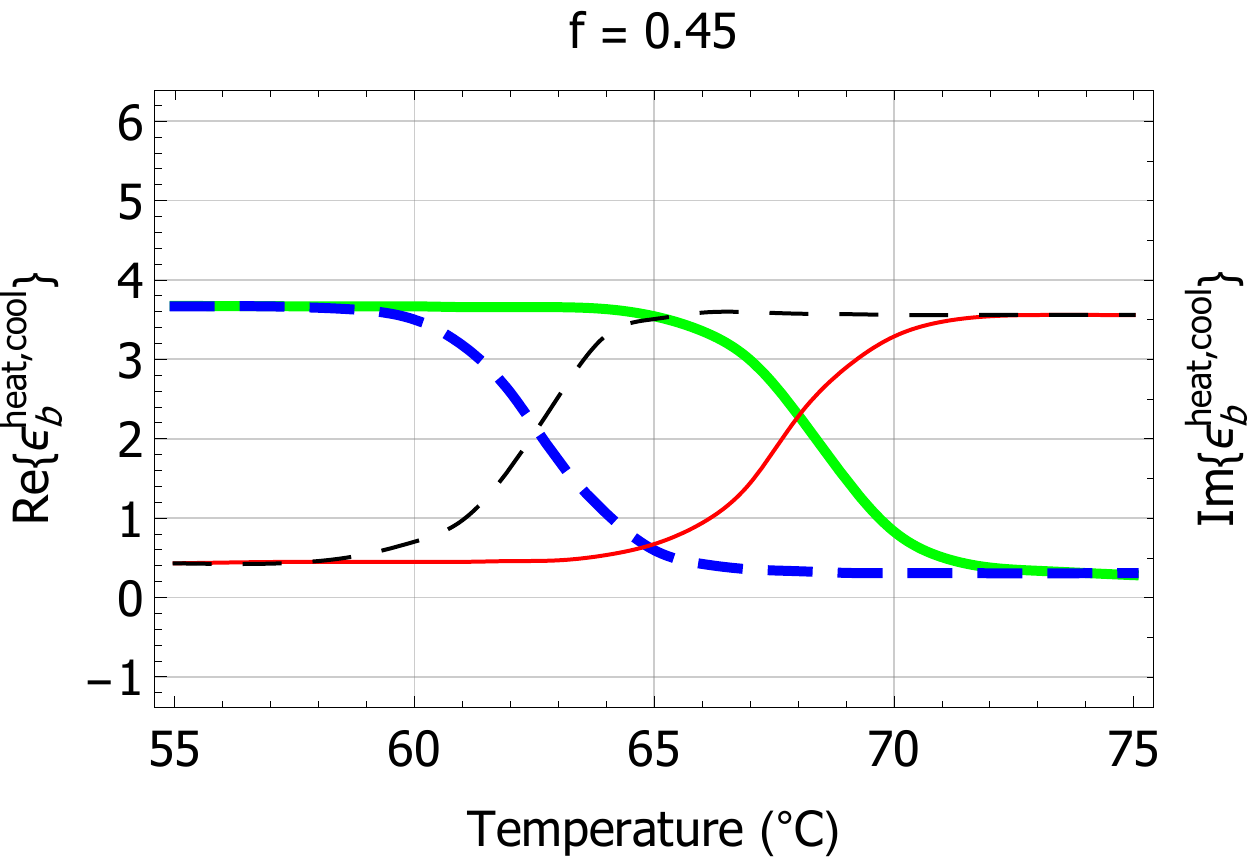}    \hspace{2mm}
\includegraphics[width=4.9cm]{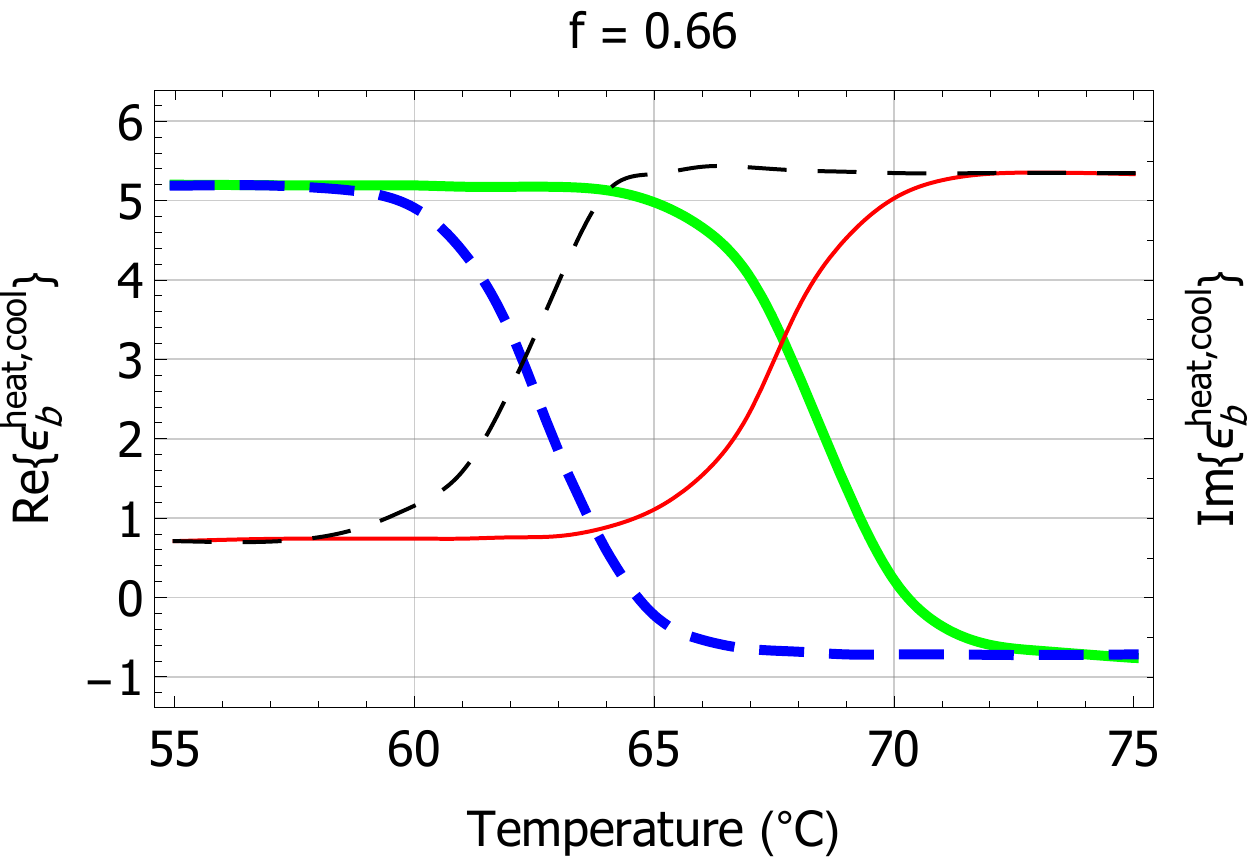}  \hspace{2mm}
\includegraphics[width=4.9cm]{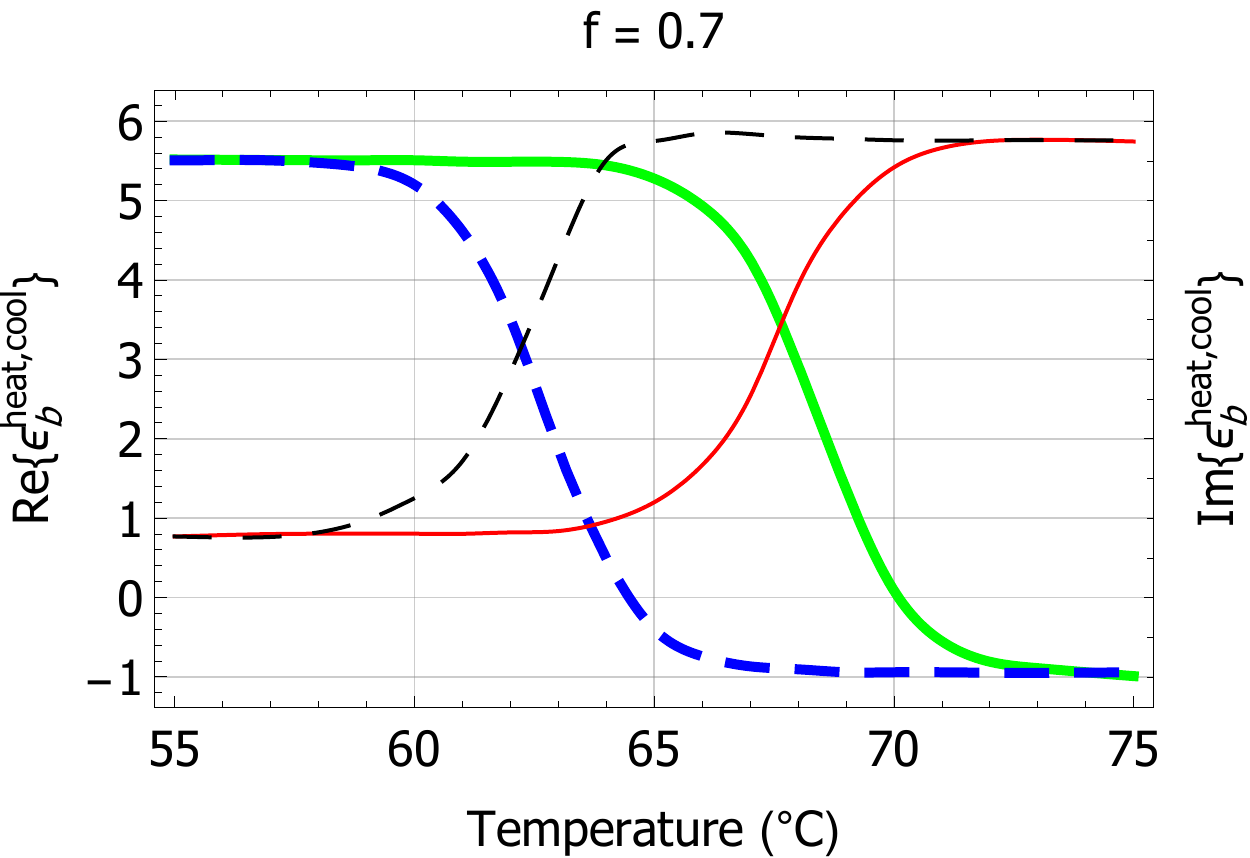}   
 \vspace{5mm}  \\
 \includegraphics[width=4.9cm]{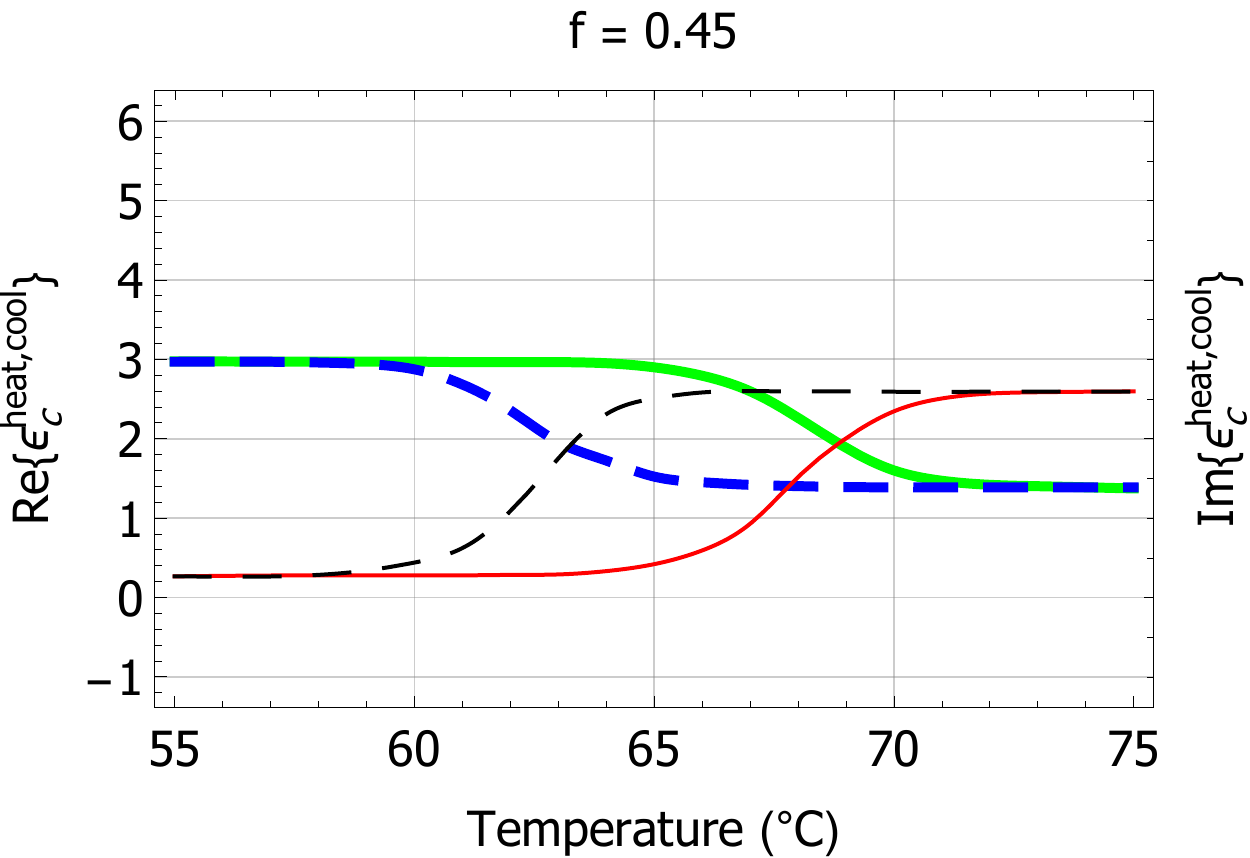}    \hspace{2mm}
\includegraphics[width=4.9cm]{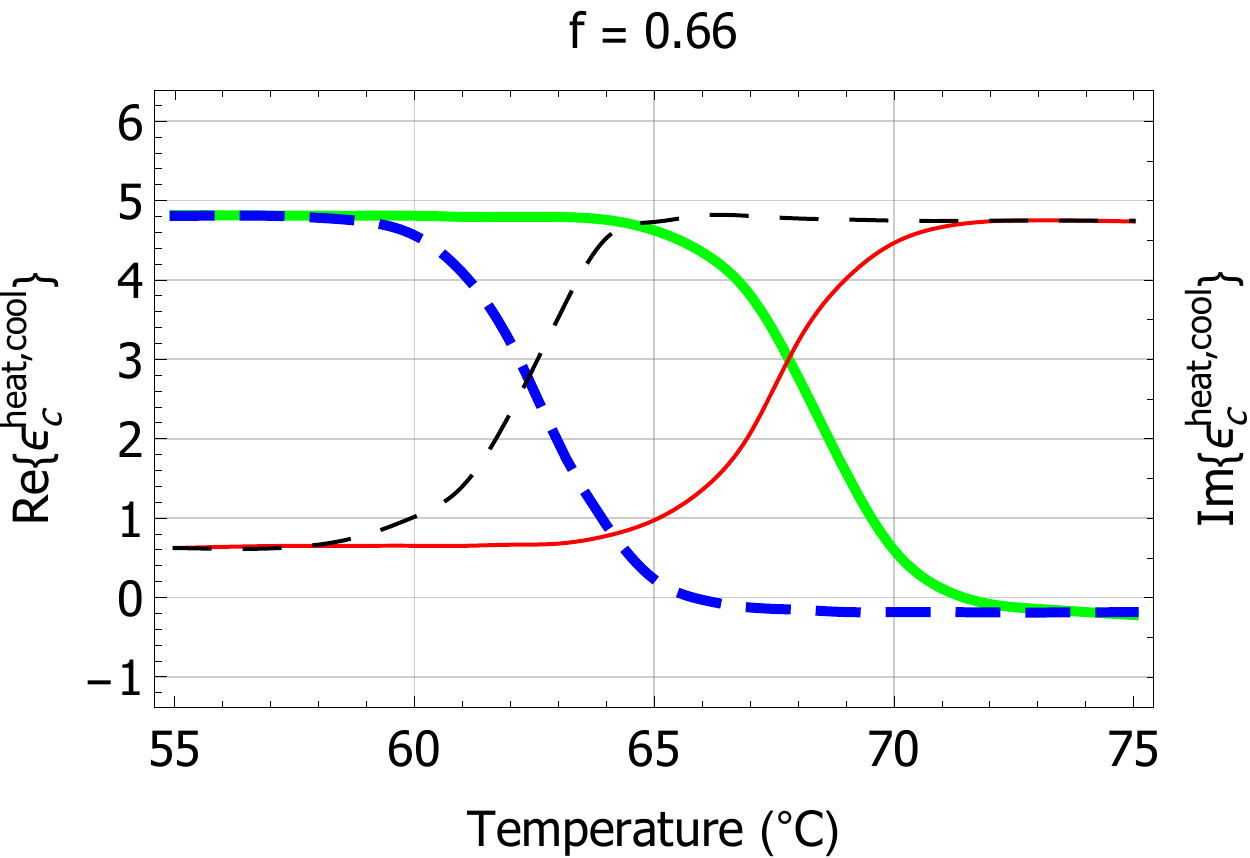}  \hspace{2mm}
\includegraphics[width=4.9cm]{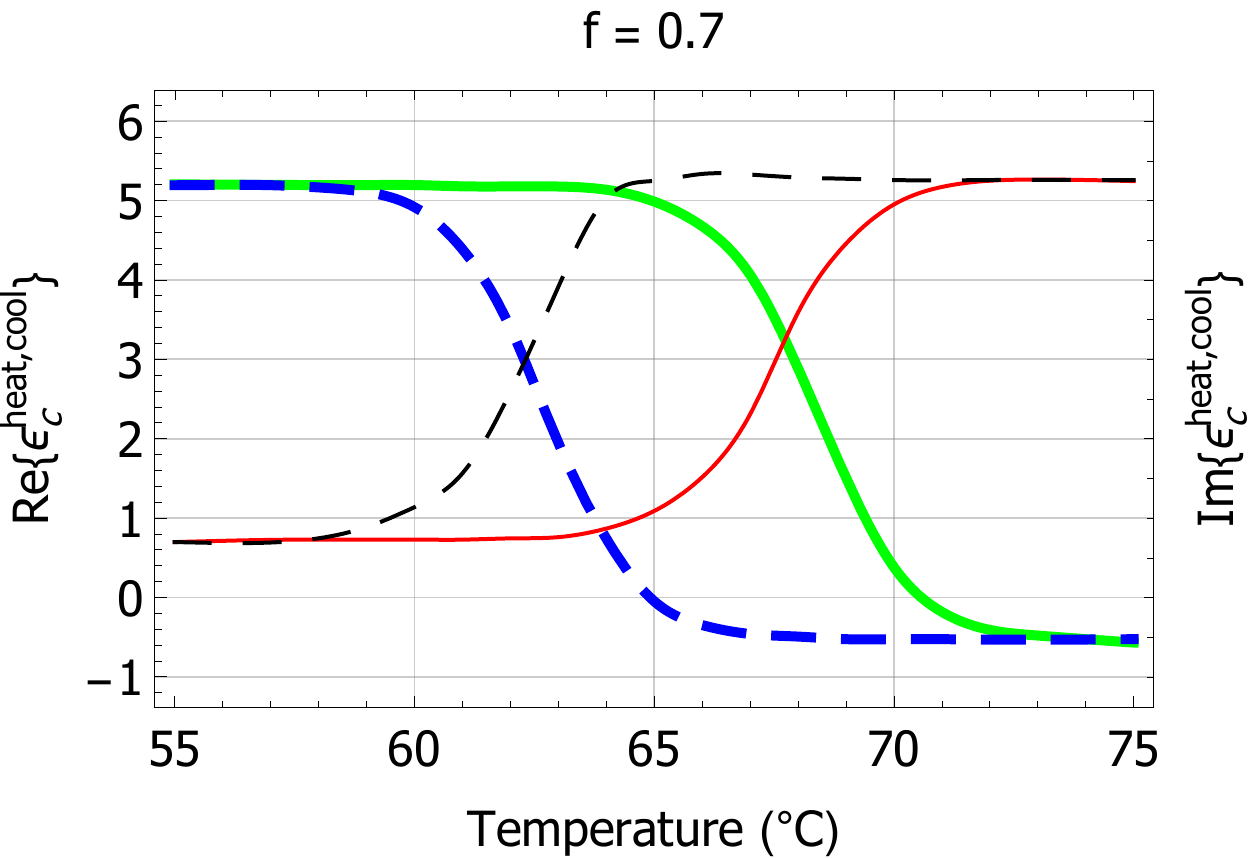} 
 \vspace{5mm}  \\
 \includegraphics[width=4.9cm]{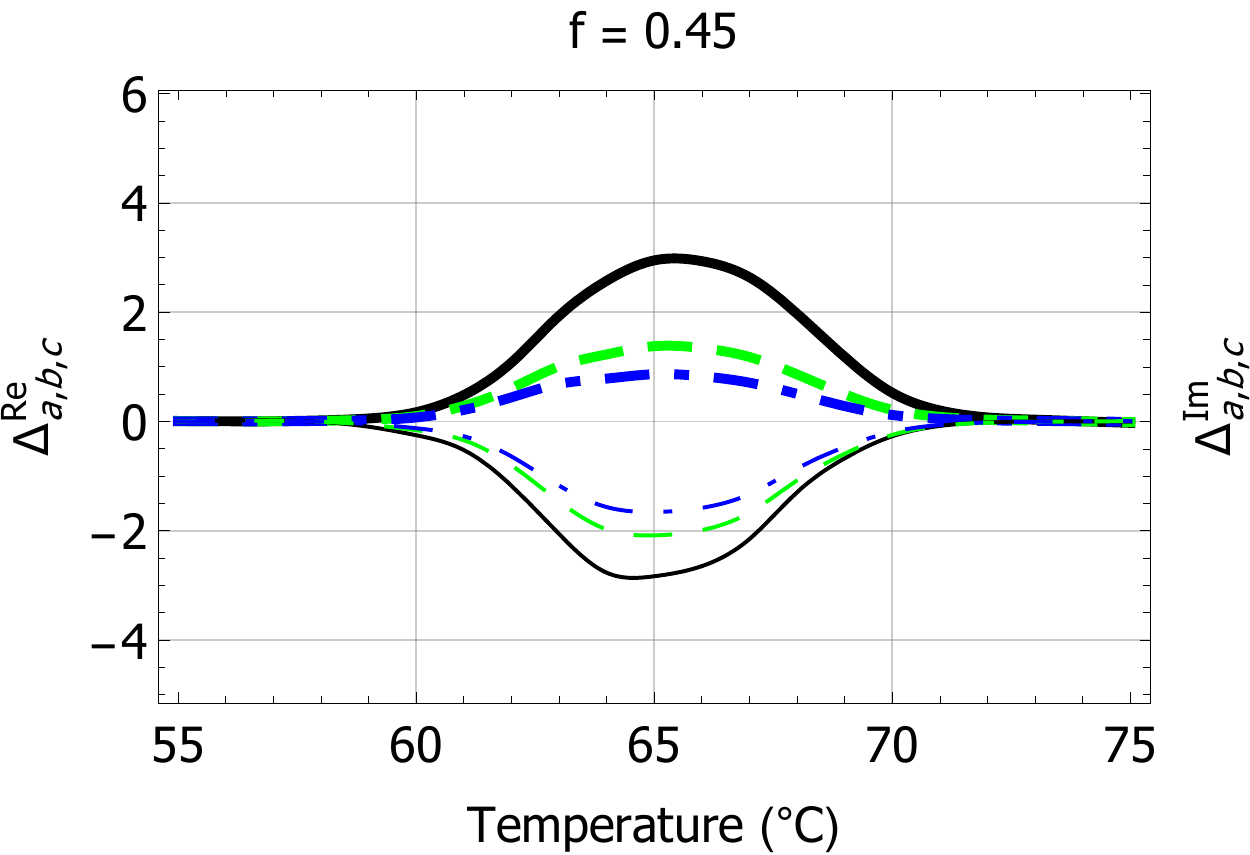}    \hspace{2mm}
\includegraphics[width=4.9cm]{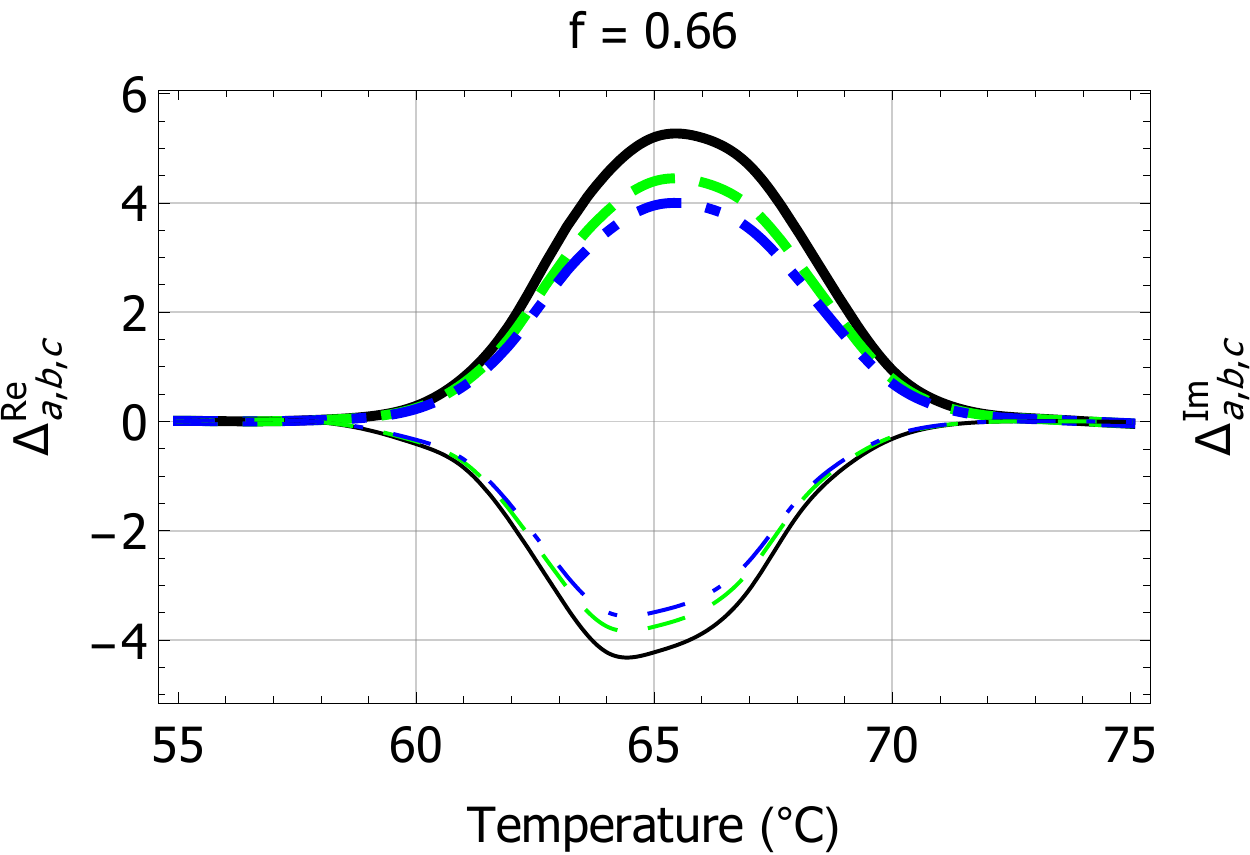}  \hspace{2mm}
\includegraphics[width=4.9cm]{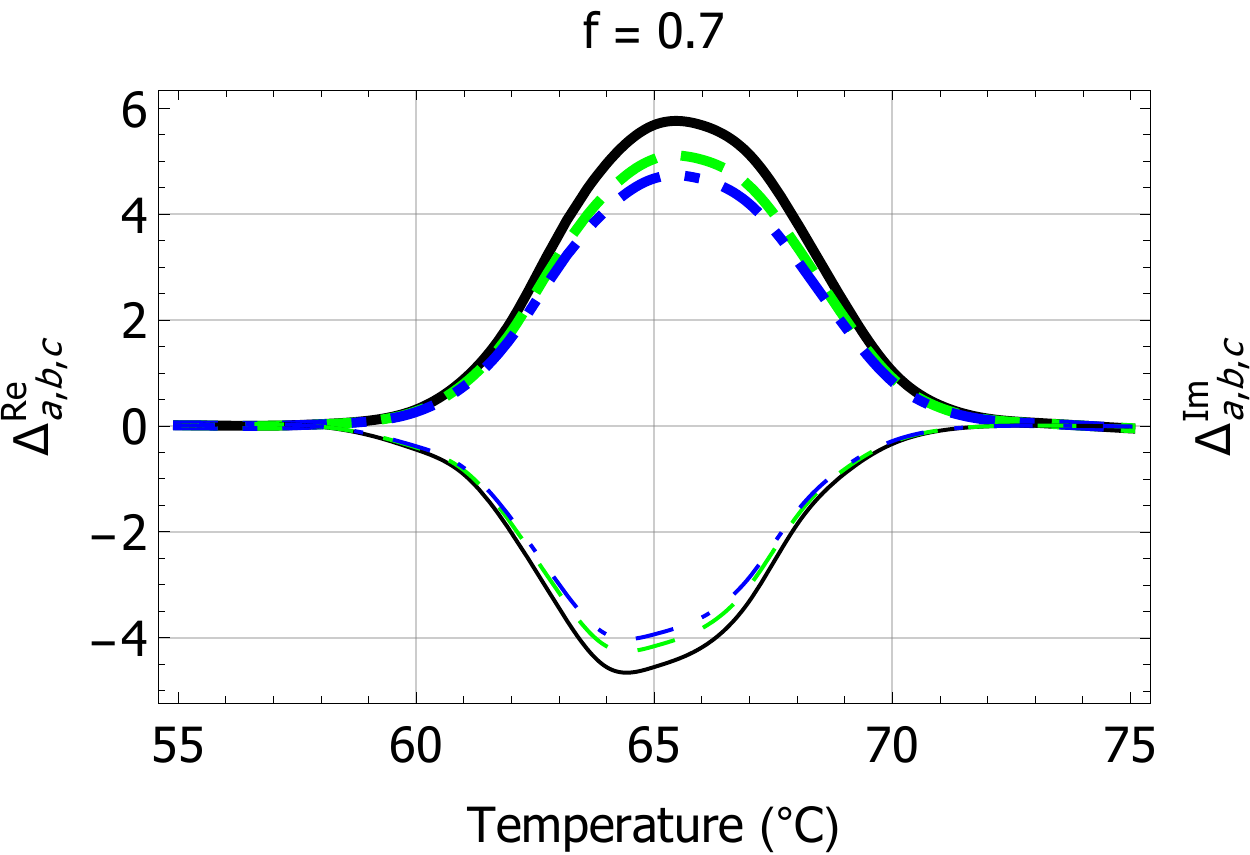}  
 \vspace{5mm}  \\
 \includegraphics[width=4.9cm]{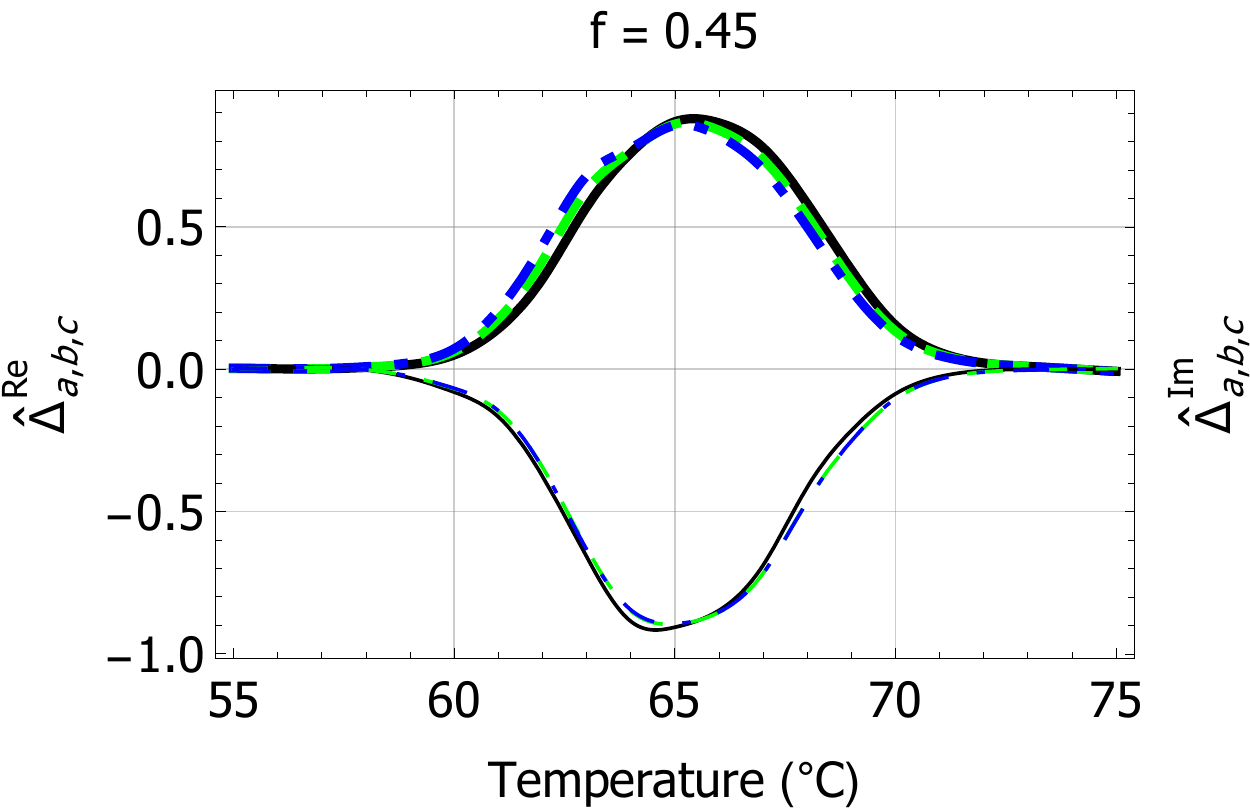}    \hspace{2mm}
\includegraphics[width=4.9cm]{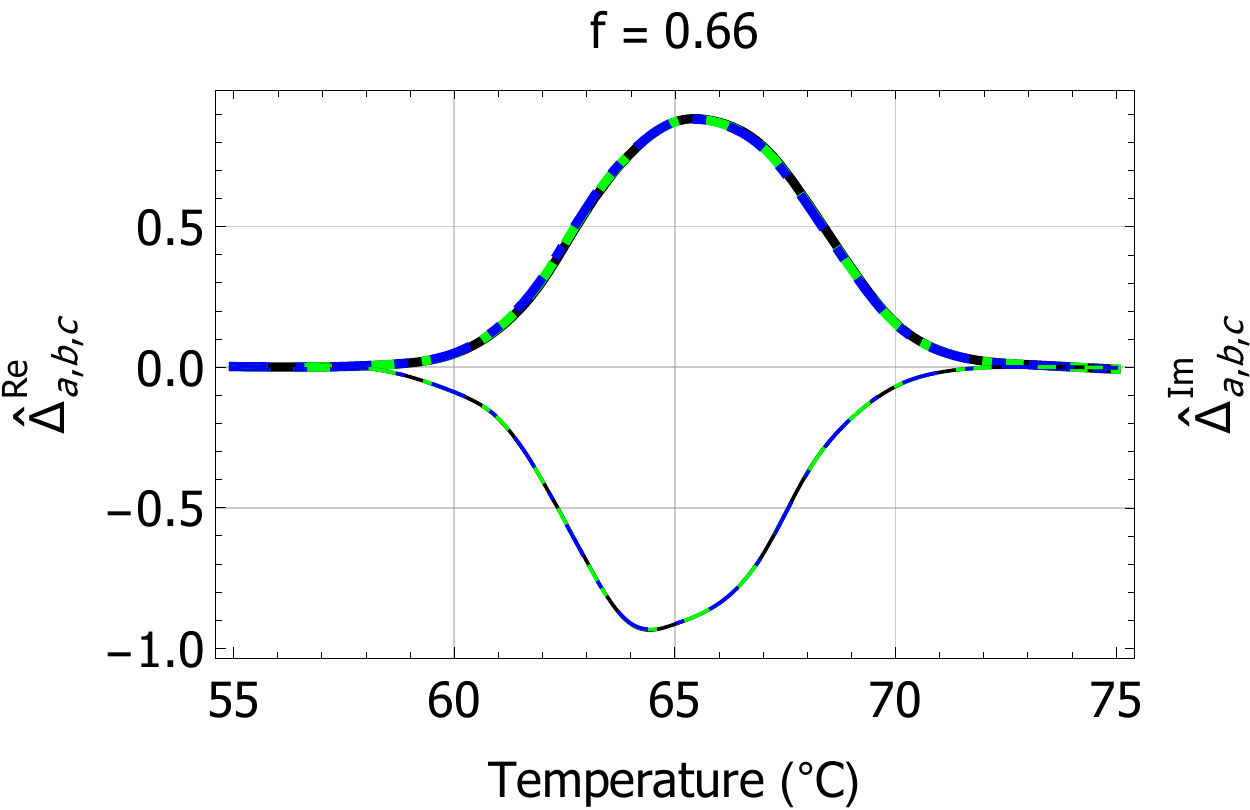}  \hspace{2mm}
\includegraphics[width=4.9cm]{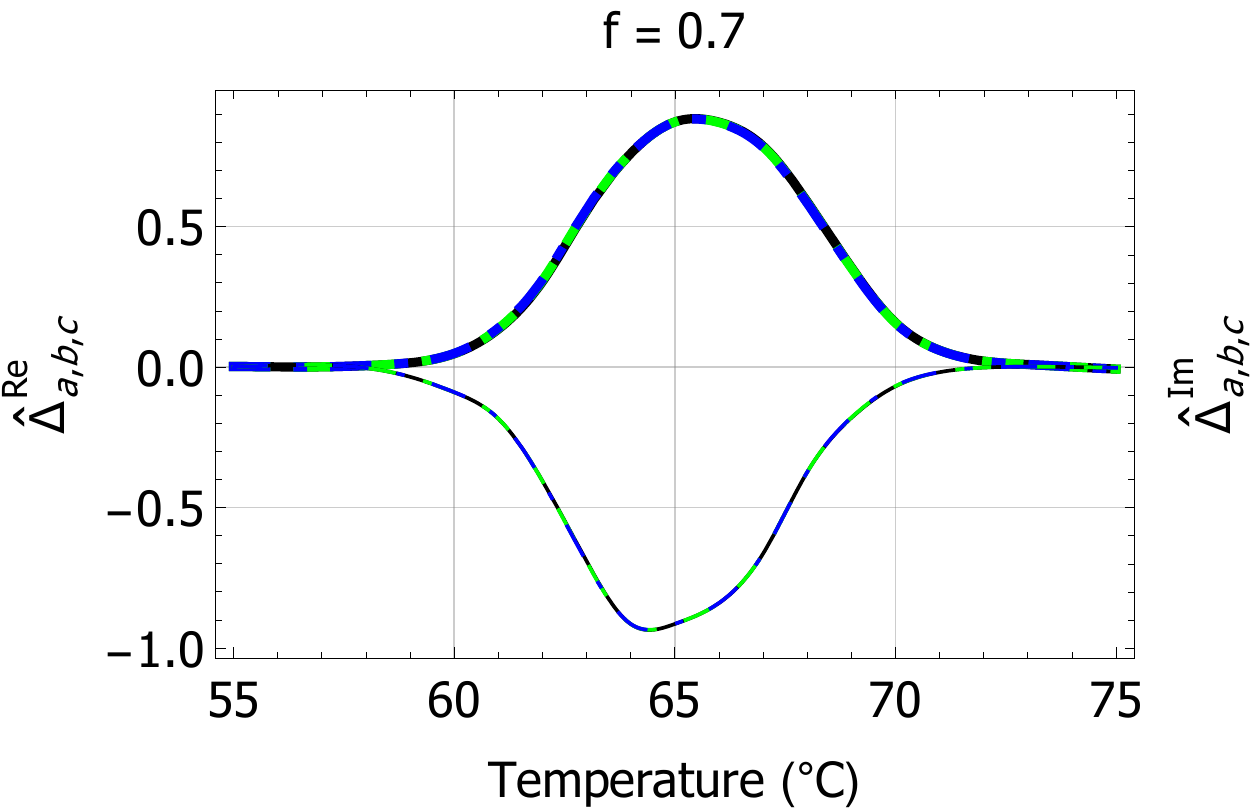}  
   \caption{\label{Fig5} As Fig.~\ref{Fig4} but for $f \in \lec 0.45, 0.66, 0.7 \ric$.
   }
\end{figure}

Now let us explore the effect of varying $f$. In order to appreciate more easily the effects of the variation  on the anisotropy of the CTF,  $\lambdao$ was fixed at 1550 nm. 
Since there can be no anisotropy for $f\in\left\{0,1\right\}$, we confined our attention to mid-range values of $f$.
The computations for Fig.~\ref{Fig4} were repeated for $f \in \lec 0.45, 0.66, 0.7 \ric$  and the results plotted in Fig.~\ref{Fig5}. 
As regards anisotropy and hysteresis, the results   for 
$f \in \lec 0.45, 0.66, 0.7 \ric$ displayed in Fig.~\ref{Fig5} are qualitatively similar to those for $f=0.6$ displayed in Fig.~\ref{Fig4}, but
certain trends are apparent:
 the anisotropy of the CTF becomes substantially more pronounced as the value of $f$ increases  
 and the anisotropy of the hysteresis,   but not of the relative hysteresis,  becomes substantially more pronounced as the value of $f$ increases. In addition, there are important differences in the natures of the CTFs for
 each of $f= 0.45$, $0.6$, $0.66$, and $0.7$:
 \begin{itemize}
\item[(i)] for $f=0.45$,  $\mbox{Re}\lec \eps_a \ric >0$, $\mbox{Re}\lec\eps_b\ric>0$, and $\mbox{Re}\lec\eps_c\ric>0$  across the entire temperature range considered;
\item[(ii)] for $f=0.6$, as discussed earlier in connection with Fig.~\ref{Fig4},  $\mbox{Re}\lec\eps_a \ric>0$ and $\mbox{Re}\lec\eps_c\ric>0$  across the entire temperature range considered
  but $\mbox{Re} \lec  \eps^{\rm{heat}}_b \ric >0 $  only for  the temperature range $55\degC$--$71\degC$   and $\mbox{Re} \lec  \eps^{\rm{cool}}_b \ric >0 $  only for  the temperature range $55\degC$--$65\degC$;
\item[(iii)] for $f=0.66$,    $\mbox{Re}\lec\eps_a \ric >0$    across the entire temperature range considered
  but $\mbox{Re} \lec  \eps^{\rm{heat}}_b \ric >0$  only for  the temperature range $55\degC$--$70.5\degC$  and $\mbox{Re} \lec  \eps^{\rm{cool}}_b \ric >0$  only for  the temperature range $55\degC$--$64\degC$;
  and
   $\mbox{Re} \lec  \eps^{\rm{heat}}_c \ric >0 $  only for  the temperature range $55\degC$--$72\degC$ 
    and $\mbox{Re} \lec  \eps^{\rm{cool}}_c \ric >0$ only for  the temperature range $55\degC$--$66\degC$;
  \item[(iv)] for $f=0.7$,
   $\mbox{Re} \lec  \eps^{\rm{heat}}_a \ric>0 $  only for  the temperature range $55\degC$--$71\degC$ 
   and $\mbox{Re} \lec  \eps^{\rm{cool}}_a \ric>0 $ only for  the temperature range $55\degC$--$65.5\degC$;
   $\mbox{Re} \lec  \eps^{\rm{heat}}_b \ric >0$  only for  the temperature range $55\degC$--$70\degC$  
    and $\mbox{Re} \lec  \eps^{\rm{cool}}_b \ric>0 $ only for  the temperature range $55\degC$--$64\degC$;
    and   $\mbox{Re} \lec  \eps^{\rm{heat}}_c \ric>0 $ only for  the temperature range $55\degC$--$70.5\degC$    and $\mbox{Re} \lec  \eps^{\rm{cool}}_c \ric >0 $ only for  the temperature range $55\degC$--$64.5\degC$.
  \end{itemize}
In other words,
the CTF is always a dissipative dielectric material at 
temperatures close to $55\degC$, for all values of $f$ considered; but
at temperatures close to $75\degC$
  the CTF can be (a) a dissipative dielectric material, (b) a hyperbolic material, or (c) a metal-like material, depending on the value of $f$.
Furthermore, the temperature at which the  dissipative dielectric material/hyperbolic material transition or the hyperbolic material/metal-like material transition occurs depends upon whether the material is being heated or cooled.

\begin{figure}[!htb]
\centering
\includegraphics[width=5.5cm]{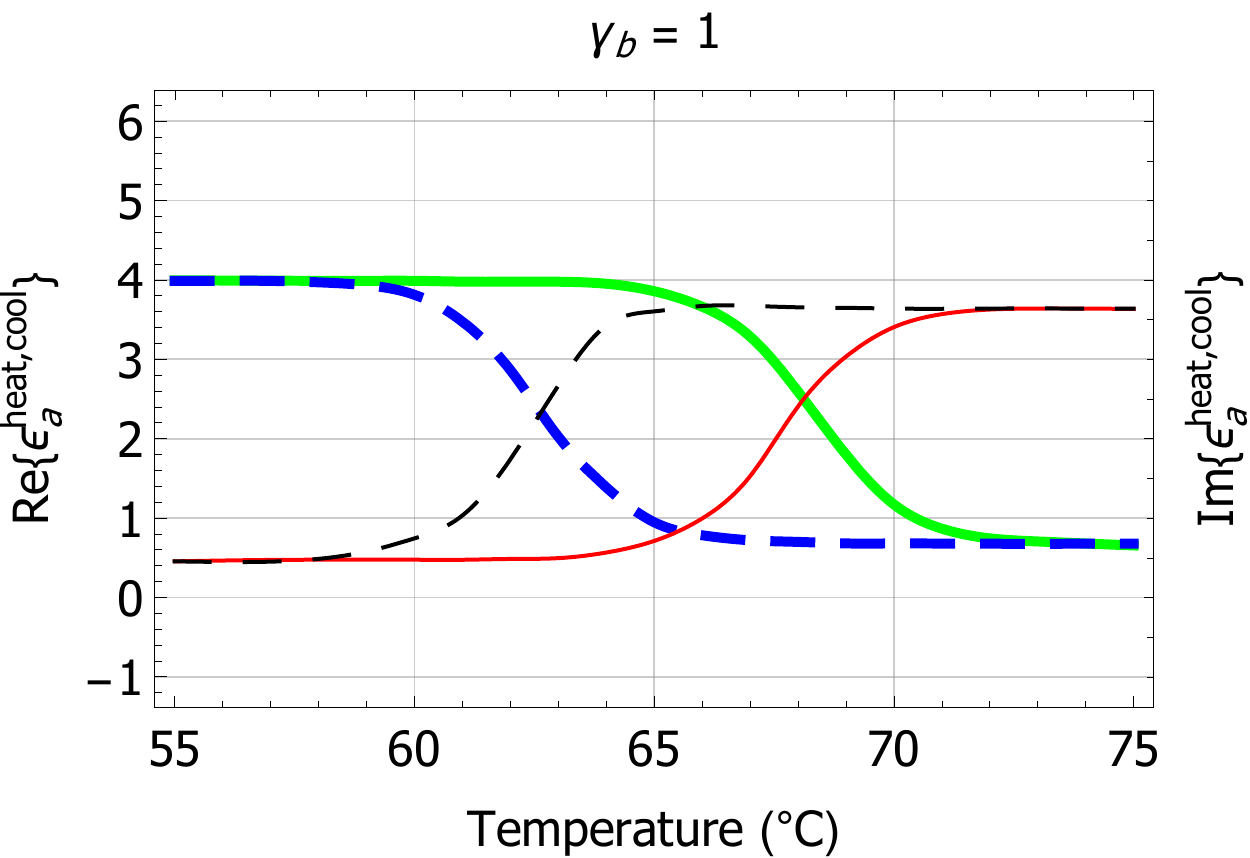}    \hspace{5mm}
\includegraphics[width=5.5cm]{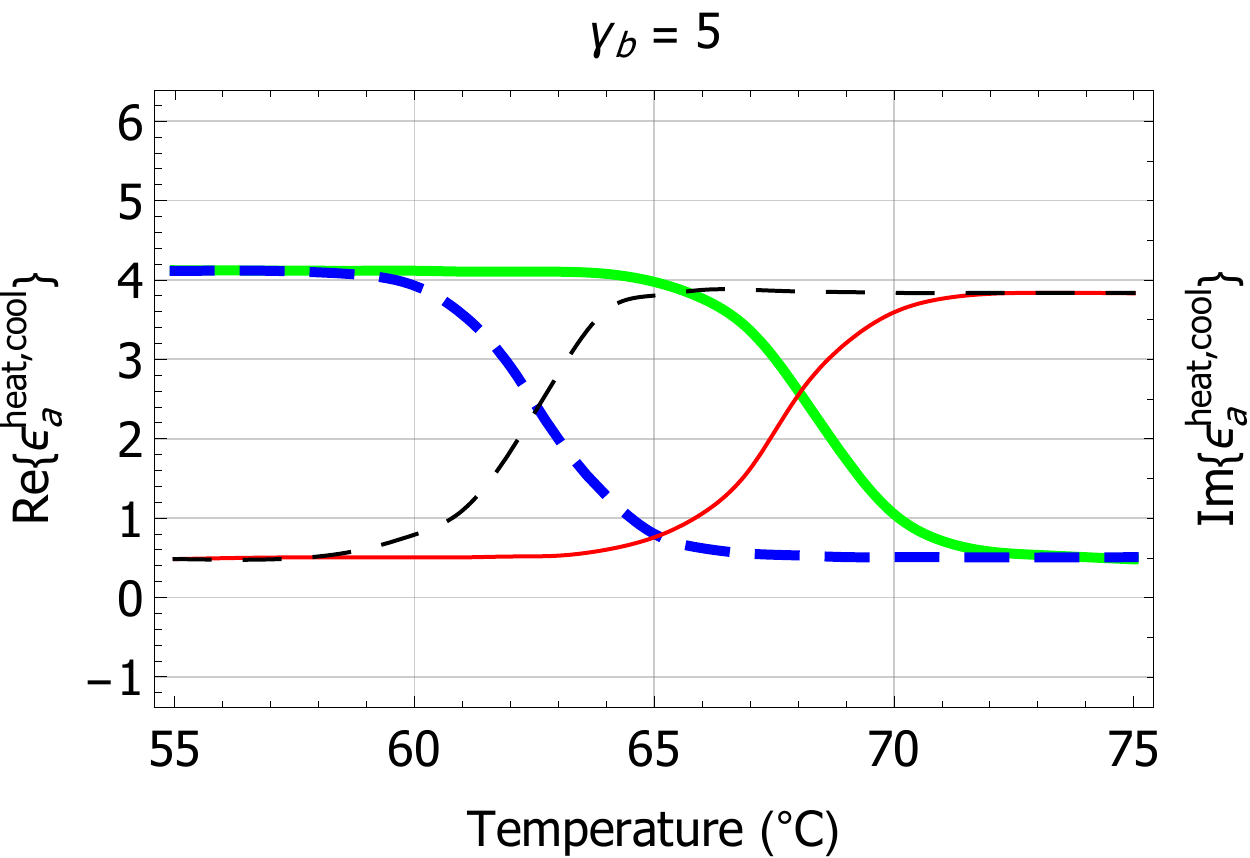}    
 \vspace{5mm}  \\
 \includegraphics[width=5.5cm]{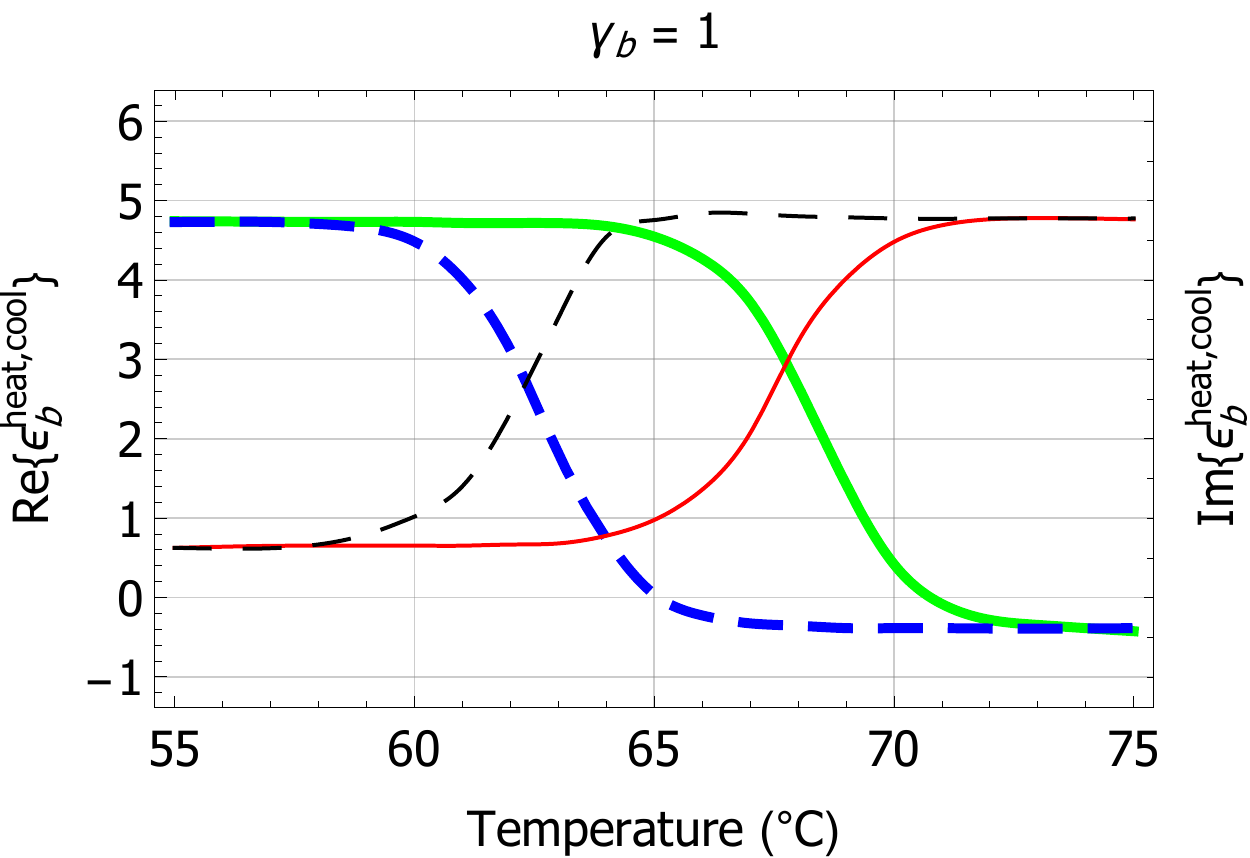}   \hspace{5mm} 
\includegraphics[width=5.5cm]{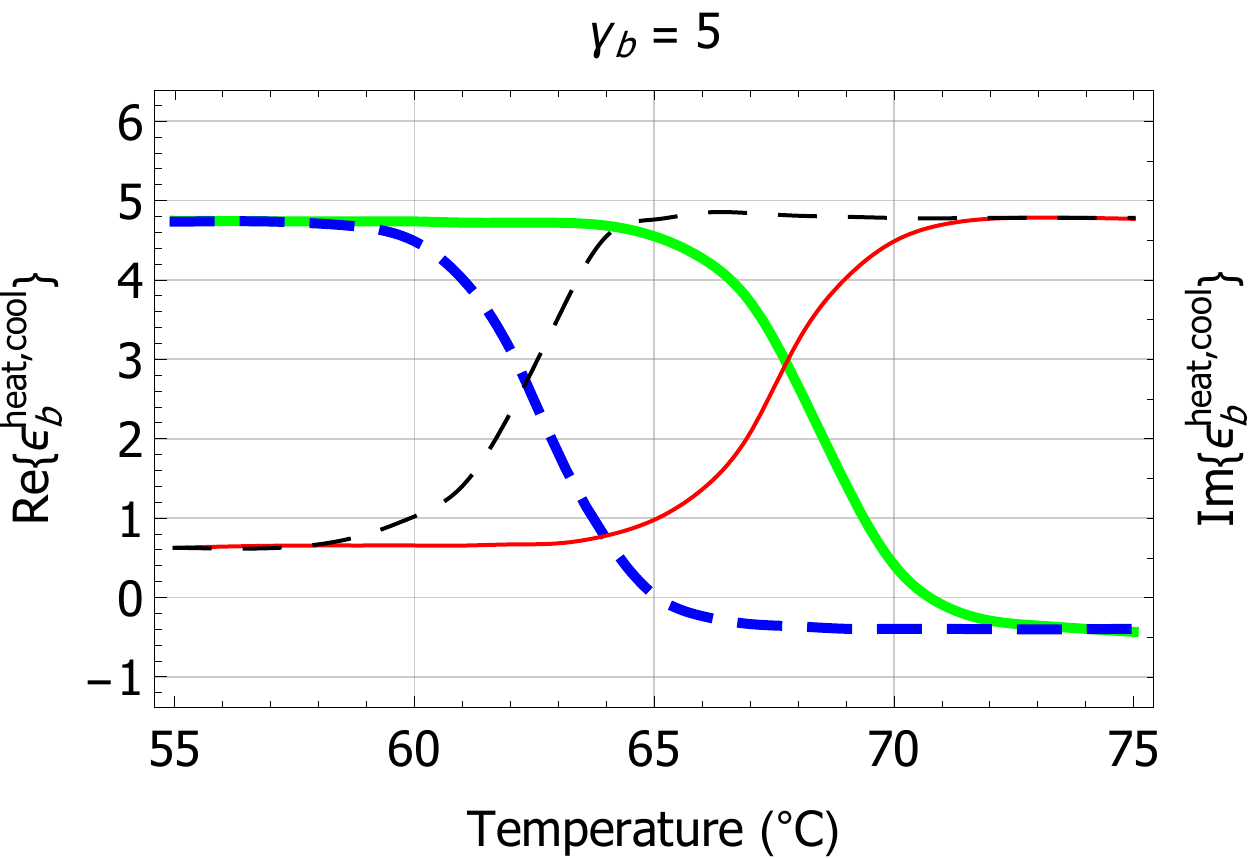}  
 \vspace{5mm}  \\
 \includegraphics[width=5.5cm]{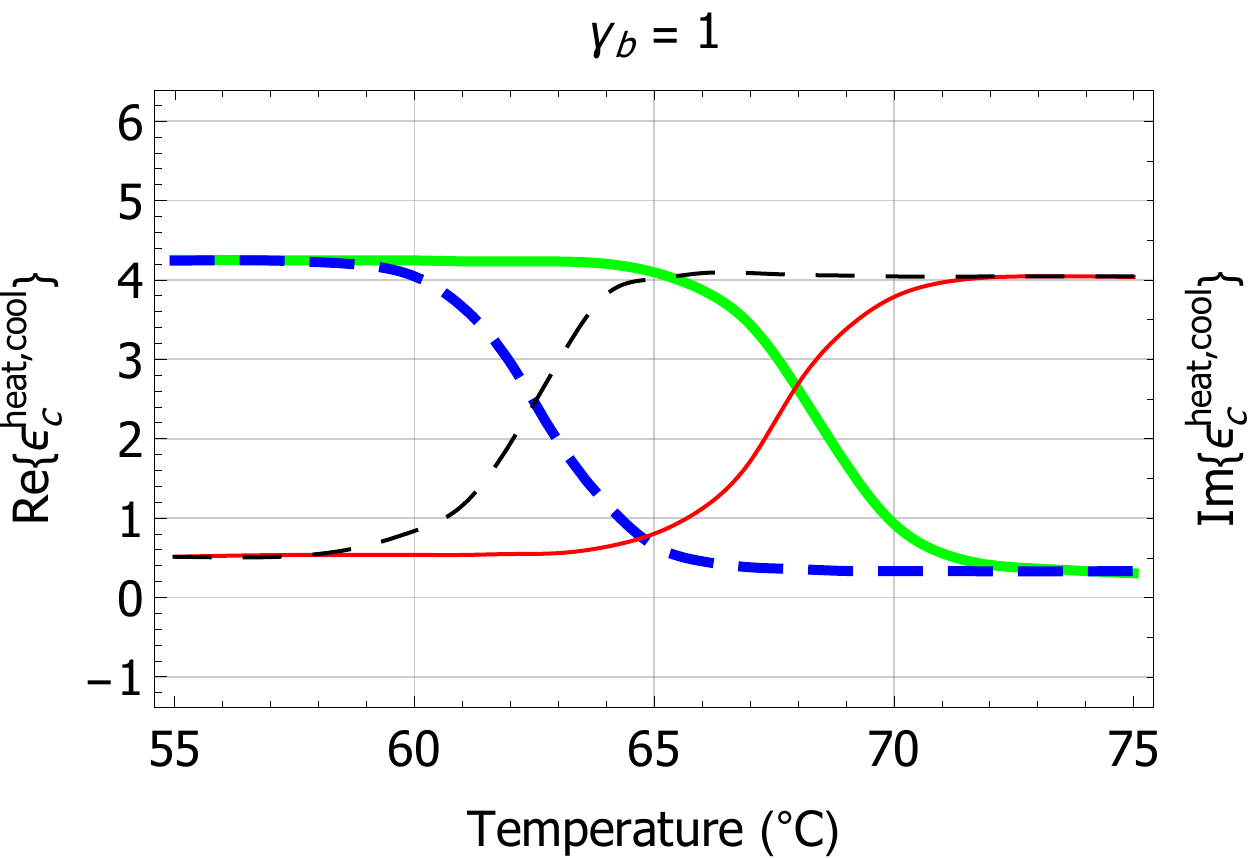}   \hspace{5mm} 
\includegraphics[width=5.5cm]{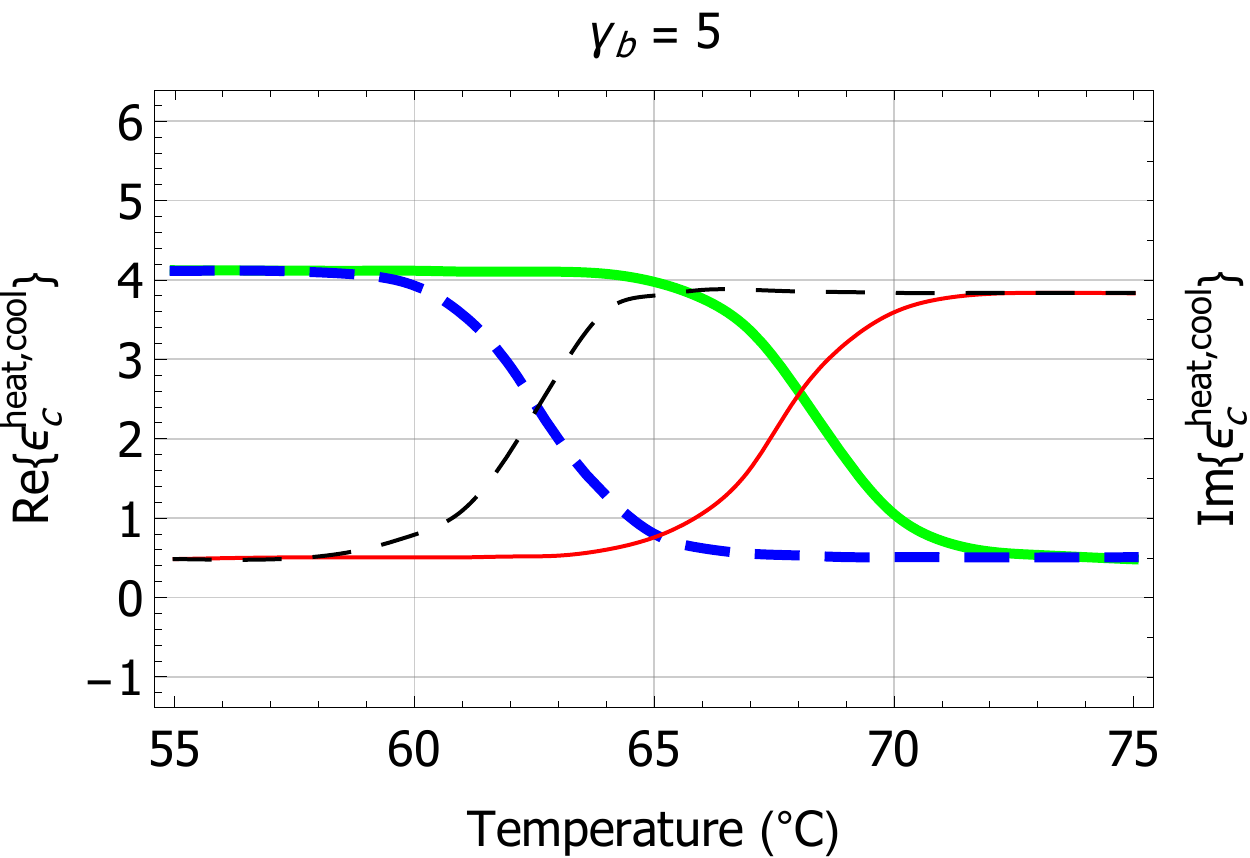}  
  \vspace{5mm}  \\
 \includegraphics[width=5.5cm]{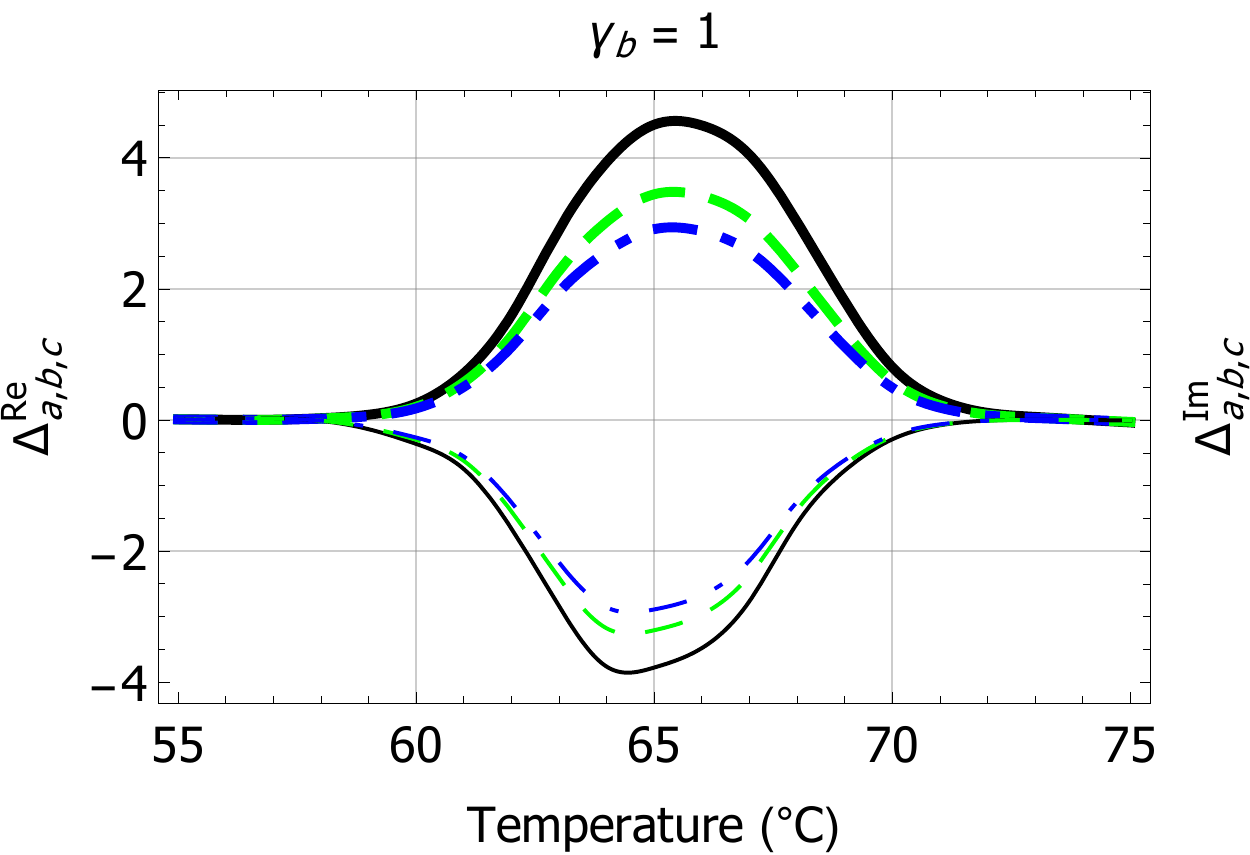}   \hspace{5mm} 
\includegraphics[width=5.5cm]{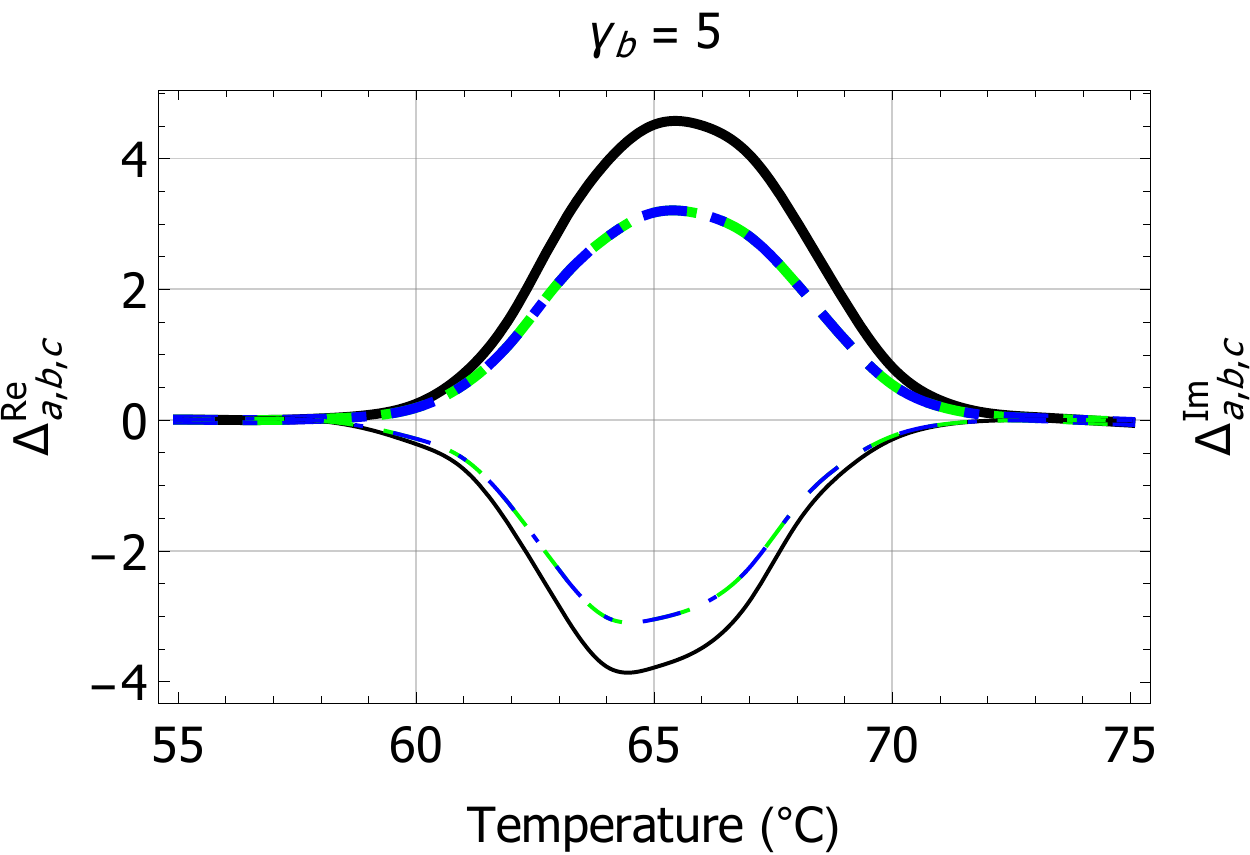}     \vspace{5mm}  \\
 \includegraphics[width=5.5cm]{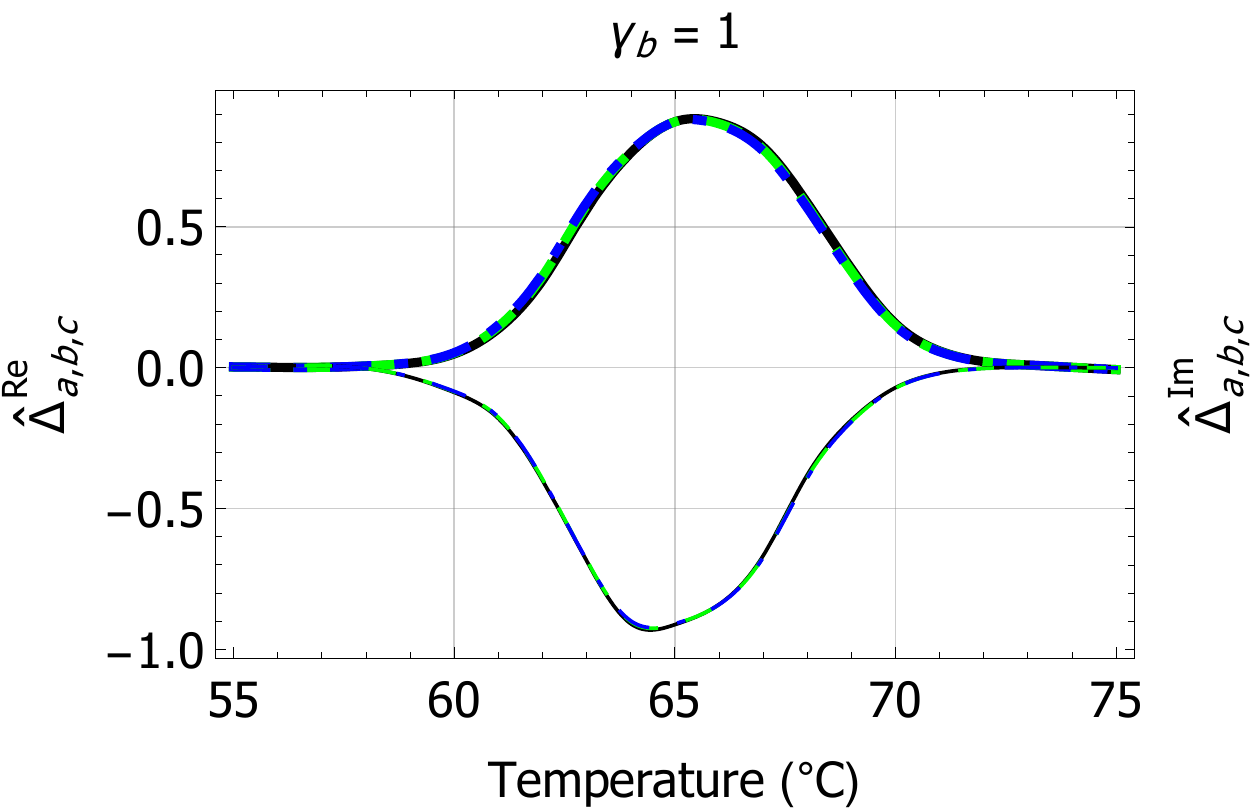}   \hspace{5mm} 
\includegraphics[width=5.5cm]{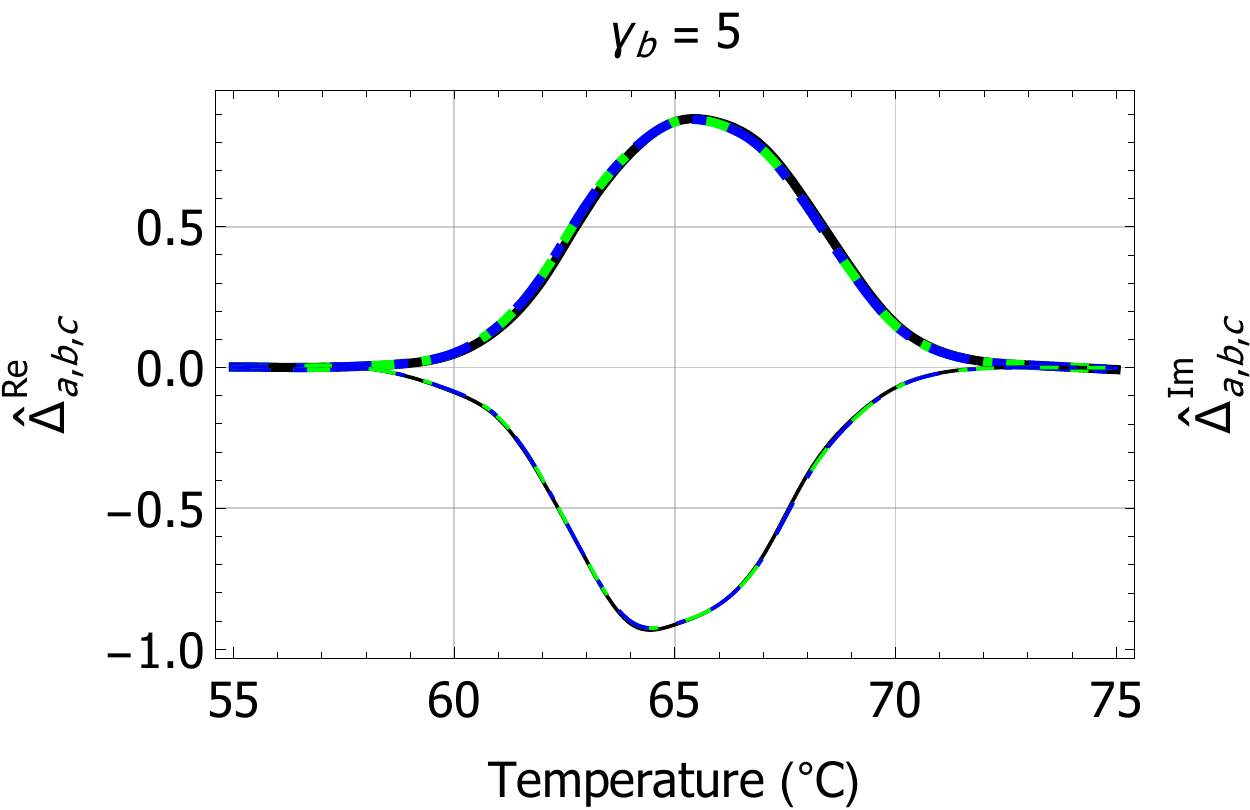}   
   \caption{\label{Fig6} As Fig.~\ref{Fig4} but for $\gamma_b \in \lec 1, 5 \ric$.
   }
\end{figure}

Lastly let us explore the effect of varying $\gamma_{\rm{b}}$. As for Fig.~\ref{Fig5},
  $\lambdao$ was fixed at 1550 nm  in order to appreciate more easily the effects of the variation on the anisotropy of the CTF. 
The computations for Fig.~\ref{Fig4} were repeated for $\gamma_{\rm{b}} \in \lec 1, 5 \ric$  and the results plotted in Fig.~\ref{Fig6}. 
The columns of the CTF are circular in cross section when $\gamma_b=1$ but like blades when $\gamma_b=5$ \cite{Tang}.
As regards anisotropy and hysteresis, the results   for 
$\gamma_{\rm{b}} \in \lec 1, 5 \ric$ displayed in Fig.~\ref{Fig6} are qualitatively similar to those for $\gamma_{\rm{b}}=1.5$ displayed in Fig.~\ref{Fig4}, but there are
certain notable features: When $\gamma_{\rm{b}} = 1$, $\eps_a = \eps_c$ for both heating and cooling phases; correspondingly, the CTF is a uniaxial material with distinguished axis parallel to $\ut$. When $\gamma_{\rm{b}}= 5$, $\eps_b \approx \eps_{\rm{c}}$ for both heating and cooling phases; correspondingly, the CTF is very close to being a uniaxial material with distinguished axis parallel to $\un$.
Only relatively modest variations in anisotropy and hysteresis arise as $\gamma_{\rm{b}}$ increases from 1 to 5, as compared to the variations induced by varying $f$.

\section{Closing remarks}

The Bruggeman homogenization formalism was used to numerically investigate the anisotropic dielectric properties of a  CTF
made from
VO${}_2$.  In the visible and near-infrared spectral regimes, the CTF is electromagnetically equivalent to a homogeneous orthorhombic material,
characterized by the relative permittivity dyadic $ \=\eps_{\,\rm{CTF}}$ with eigenvalues $\lec \eps_a, \eps_b, \eps_c \ric$.
Over the temperature range $58\degC$--$72\degC$,
$\lec \eps_a, \eps_b, \eps_c \ric$ are highly sensitive to temperature, and vary according to whether the CTF is being heated or cooled.
 The  anisotropy revealed through $\lec \eps_a, \eps_b, \eps_c \ric$, and the anisotropy of the associated hysteresis, were investigated  as a function of temperature for
  CTFs of different porosities and of different column shapes.
The CTF is a dissipative dielectric material that exhibits temperature-dependent anisotropy and anisotropic hysteresis,
for $\lambdao= 800$~nm. In contrast, for
$\lambdao=1550$~nm, the CTF can be either a dissipative dielectric material, a hyperbolic material or a metal-like material, depending on the temperature and
  on the porosity of the CTF.
  
  The ability to modulate the anisotropy of a CTF made from
VO${}_2$ with temperature is an attractive feature  when considering directional control and tuning for applications involving reconfigurable and multifunctional devices.
Furthermore, the ability to modulate anisotropic thermal hysteresis may be an attractive feature for directional control in optical/infrared switching applications. And the prospect of
thermally controlling
 whether a 
 CTF is a dissipative dielectric material, a hyperbolic material or a metal-like material should present additional opportunities \cite{Takayama,Davidovich}.

 \vspace{10mm}
 
 \noindent {\bf Acknowledgments:} 
TGM was supported  by
EPSRC (grant number EP/V046322/1).  AL   was supported by the US National Science Foundation (grant number DMS-1619901)  as well as   the Charles Godfrey Binder Endowment at Penn State.\\


\newpage

\newpage

\newpage

\newpage

\newpage

\newpage

\end{document}